\documentclass[%
superscriptaddress,
twocolumn,
amsmath,
amssymb,
aps,
prb,
floatfix,
]{revtex4-2}

\usepackage{graphicx}
\usepackage[colorlinks]{hyperref}
\hypersetup{%
	plainpages=true,
	breaklinks=true,
	hypertexnames=false,
	pageanchor=true,
	colorlinks=true,
	linkcolor={blue},
	citecolor={red},
	urlcolor={blue},
	anchorcolor={black}
}

\usepackage[dvipsnames]{xcolor}
\usepackage[normalem]{ulem}
\usepackage{amsthm}
\usepackage{mathtools}

\begin{document}
\title{
Linear response theory for cavity QED materials at arbitrary light-matter coupling strengths
}

\author{Juan Román-Roche}
\affiliation {Instituto de Nanociencia y Materiales de Aragón (INMA), CSIC-Universidad de Zaragoza, Zaragoza 50009, Spain}
\affiliation{Departamento de Física de la Materia Condensada, Universidad de Zaragoza, Zaragoza 50009, Spain}

\author{Álvaro Gómez-León}
\affiliation {Institute of Fundamental Physics IFF-CSIC, Calle Serrano 113b, 28006 Madrid, Spain}

\author{Fernando Luis}
\affiliation {Instituto de Nanociencia y Materiales de Aragón (INMA), CSIC-Universidad de Zaragoza, Zaragoza 50009, Spain}
\affiliation{Departamento de Física de la Materia Condensada, Universidad de Zaragoza, Zaragoza 50009, Spain}

\author{David Zueco}
\affiliation {Instituto de Nanociencia y Materiales de Aragón (INMA), CSIC-Universidad de Zaragoza, Zaragoza 50009, Spain}
\affiliation{Departamento de Física de la Materia Condensada, Universidad de Zaragoza, Zaragoza 50009, Spain}
  
\date{\today}

\begin{abstract}
We develop a linear response theory for materials collectively coupled to a cavity that is valid in all regimes of light-matter coupling, including symmetry-broken phases. We present and compare two different approaches. First, using a coherent path integral formulation for the partition function to obtain thermal Green functions. This approach relies on a saddle point expansion for the action, that can be truncated in the thermodynamic limit. Second, by formulating the equations of motion for the retarded Green functions and solving them. We use a mean-field decoupling of high-order Green functions in order to obtain a closed, solvable system of equations. Both approaches yield identical results in the calculation of response functions for the cavity and material. These are obtained in terms of the bare cavity and material responses. In combination, the two techniques clarify the validity of a mean-field decoupling in correlated light-matter systems and provide complementary means to compute finite-size corrections to the thermodynamic limit. The theory is formulated for a general model that encompasses most of the systems typically considered in the field of cavity QED materials. Finally, we provide a detailed application of the theory to the Quantum Hall effect and to a collection of spin models. 
\end{abstract}

\maketitle

\section{Introduction}
\label{intro}

Linear Response Theory (LRT) is an elegant tool for understanding how systems at equilibrium respond when they are perturbed. The  linearity refers to the condition that the perturbation is sufficiently weak, allowing the system's dynamics to be treated to first order in the perturbation. This framework yields various relations that determine response properties, regardless of the specific situation or model considered. Fundamentally, it relates equilibrium fluctuations and dissipation through the fluctuation-dissipation theorem \cite{altland2010condensed}[Chap. 7]. 

In this paper, we investigate a particular scenario where a system containing $N$ local degrees of freedom is coupled to a single-mode cavity. The coupling is assumed to be collective, and our theory yields exact results in the thermodynamic limit for the matter, $ N \to \infty $. This is the typical limit considered in cavity QED materials, on which we will elaborate below \cite{schlawin2022cavity, garciavidal2021manipulating, bloch2022strongly}. Beyond the assumption of collective coupling and the thermodynamic limit, our approach is general. We obtain formulas for the response of both the material and the cavity in terms of the bare photonic and material response functions, for any light-matter coupling strength.
At equilibrium, it has been shown that the \( N \to \infty \) limit simplifies the problem, allowing for exact results, particularly famous is the existence (or lack thereof)  of the superradiant phase transition. Various techniques have been employed, including thermodynamic inequalities \cite{hepp1973on}, scaling arguments \cite{wang1973phase}, the stiffness theorem \cite{andolina2019cavity}, and path integral methods \cite{romanroche2022effective}. In this work, we are interested in perturbing the equilibrium to build an equally rigorous result for LRT. To achieve this, we use two independent but complementary approaches: one based on a path integral formulation of the partition function and the other on the equations of motion for the correlators. 
The main idea underpinning our derivation is the fact that collective interactions can be exactly decoupled by an auxiliary mean-field variable. This is a common resource in large-$N$ theories in quantum field theory which is used in combination with a saddle-point approximation for the action \cite{moshe2003quantum}. The complementary formulation in terms of the equations of motion for the correlators renounces the provable exactness afforded by the path integral in exchange for a simpler derivation where the mean-field decoupling is explicit.

Cavity QED materials are actively discussed in the literature, particularly the possibility of modifying the properties of matter using quantum light \cite{schlawin2022cavity, garciavidal2021manipulating,bloch2022strongly}. This is pertinent in the regime of strong light-matter correlations, which is why a non-perturbative approach is relevant.  The problem is far from being a theoretical curiosity, as initial experiments have reported modifications in conductivity \cite{paravicinibagliani2018magnetotransport, appugliese2022breakdown}, magnetism \cite{thomas2021large}, and metal-to-insulator transition \cite{jarc2023cavity}. Remarkably, one way to observe these modifications is through the hybrid light-matter response \cite{flick2019lightmatter}, which is where our work fits, providing a general theory for such situations.
In this context, our work presents the LRT for cavity materials using both a path-integral and an equations-of-motion approach. We discuss their relationship and the application of our theory. We demonstrate that our theory recovers previous results on the cavity-modified quantum Hall effect and explore the response of magnetic materials in cavities, discussing dipolar Ising systems with longitudinal and transverse coupling to the cavity and Heisenberg ferromagnetism. The theory has also been applied to the Dicke-Ising model \cite{romanroche2024linear}.

The paper is organized as follows. In Sec. \ref{model} we present the Hamiltonian and discuss its validity to describe the typical models of cavity QED materials. In Sec. \ref{LRT} we develop and compare two distinct linear response theories. We provide expressions for different response functions of the hybrid light-matter system in terms of the bare response functions of the cavity and the material. Section \ref{qhe} is dedicated to the application of the LRT to the quantum Hall effect. More precisely to a 2D electron gas subject a homogeneous perpendicular classical magnetic field. In Sec. \ref{spinmodels} we apply the theory to several magnetic models: the Dicke model, the longitudinal and transverse Dicke-LMG models and the Heisenberg model. We end the paper with some conclusions and relegate some technical details to the appendices. 

\section{Models in cavity QED materials}
\label{model}

We will consider a Hamiltonian of the form
\begin{equation}
    H = H_{\rm m} + \Omega a^\dagger a + g \left(a + a^\dagger\right) C_{x} + \zeta \frac{g^2}{\Omega} C_{x}^2 \,. 
    \label{eq:Hstart}
\end{equation}
Here $a$ and $a^\dagger$ are bosonic annihilation and creation operators, $[a, a^\dagger]=1$, with $\Omega$ the frequency of the cavity mode [See App. \ref{app:multimode} for the multimode case]. The bare material system is described by the Hamiltonian $H_{\rm m}$, which remains unspecified beyond the fact that it describes $N$ particles or degrees of freedom. Likewise $C_{x}$ is an unspecified matter coupling operator that is Hermitian and collective, i.e. $C_x = \sum_j^N o_j$ with $o_j$ local hermitian operators of the material. The light matter coupling constant $g = \lambda/\sqrt{N}$ represents the coupling per particle, with $\lambda$ the collective coupling. The last term is a $P^2$ term, toggled by $\zeta =0, 1$, that will have to be included if the model is to describe electric dipoles coupled to the electromagnetic field of a cavity in the dipole gauge. 

Depending on the choice of $H_{\rm m}$, $C_{x}$, $\Omega$, $g$ and $\zeta$ this Hamiltonian can describe several distinct microscopic systems in different gauges \cite{bernardis2018breakdown,rokaj2018lightmatter,kockum2019ultrastrong,stefano2019resolution,stokes2022implications}. The Pauli-Fierz Hamiltonian in the Coulomb gauge can be brought to the form of Hamiltonian \eqref{eq:Hstart} by invoking the long-wavelength approximation and diagonalizing the photonic sector with a Bogoliubov transformation to eliminate the $A^2$ term \cite{rokaj2022free, andolina2019cavity, romanroche2022effective}. It would correspond to setting $\zeta = 0$ and associating the electronic collective momenta to $C_x$ and renormalized frequency and couplings, stemming from the Bogoliubov transform, to $\Omega$ and $g$ (See Sec. \ref{qhe} for an example). Fermi-Hubbard models cannot be cast to the form of Hamiltonian \eqref{eq:Hstart} due to the non-linear nature of the Peierls coupling, unless it is linearized by considering the weak coupling limit or by moving to the dipole gauge \cite{li2020electromagnetic, arwas2023quantum, dmytruk2021gauge}. Localized electric dipoles can be described in the dipole gauge by setting $\zeta = 1$ and associating the collective dipole operator to $C_x$ \cite{schuler2020thevacua, romanroche2022effective}. Since the inclusion of the $P^2$ term to ensure gauge invariance in dipolar systems is not always respected, toggling off $\zeta$ will clarify its effects within linear response theory. Furthermore, we consider $\zeta$ a binary parameter here, but promoting it to take arbitrary values also allows to study a model where the $P^2$ term amplitude is not fixed by the light-matter coupling and cavity frequency. To describe magnetic dipoles instead, one can associate a collective spin operator to $C_x$ and toggle off the $P^2$ term by setting $\zeta = 0$ \cite{libersky2021direct, mckenzie2022theory, romanroche2021photon}. Finally, a system of coupled oscillators is also compatible with Hamiltonian \eqref{eq:Hstart}, although solving it with the theory described in this paper would be overkill.

\section{Linear response theory}
\label{LRT}

Throughout this text, we will compute linear response functions, i.e. retarded Green functions, for the light and matter subsystems.
In our derivation, we will not specify the matter subsystem; instead, we will derive relationships between the Green functions for the hybrid light-matter system with the bare matter and photonic Green functions. The bare photonic Green function is a well know quantity that is trivial to compute. A full solution will thus be contingent on being able to compute the Green function of the bare matter, possibly subject to some additional mean fields, as we will gather from the derivation. The retarded Green function for operators $A$ and $B$ is defined as
\begin{equation}
    G^{\rm r}_{A, B}(t, t') = -i\theta(t - t') \langle [A(t), B(t')]\rangle \,,
    \label{eq:retardedgreenfunction}
\end{equation}
With $\theta(t)$ the Heaviside step function.
We will be particularly concerned with the photonic propagator
\begin{equation}
    D(t) = G^{\rm r}_{a, a^\dagger}(t, 0) 
\end{equation}
and matter response functions
\begin{equation}
    \chi_{rs}(t) = -\frac{1}{N} G^{\rm r}_{C_{r}, C_{s}}(t, 0) \,,
\end{equation}
where $C_r$ might be $C_x$ or any other collective matter operator.

\subsection{Imaginary-time path integral: Thermal Green function}

One way to compute retarded Green functions is to obtain them by analytic continuation from the thermal Green function, which can be computed from an imaginary-time path integral formulation of the partition function of the system. The thermal Green function for operators $A$ and $B$ reads
\begin{equation}
    G^{\rm t}_{A, B} (\tau - \tau') = -\langle \mathcal T_\tau A(\tau) B(\tau') \rangle
    \label{eq:thermalgreen}
\end{equation}
Within the imaginary-time formalism, we will obtain the thermal Green functions in terms of imaginary time, $\tau$, or Matsubara frequencies, $\omega_m$. The retarded Green function is then obtained with the replacement 
\begin{equation}
    G^{\rm r}_{A, B} (\omega) = G^{\rm t}_{A, B} (\omega_m) |_{i\omega_m \to \omega_+ = \omega + i 0^+} \; ,
\end{equation}
for all positive Matsubara frequencies \cite{altland2010condensed}[Chap. 7].

\subsubsection{Effective action}
We formulate the partition function as a path integral over coherent states
\begin{equation}
    Z = \oint_{a, \bar a, c}  e^{-S} \,,
    \label{eq:ogaction}
\end{equation}
with
\begin{equation}
\begin{split}
    S =&  S_{\rm m} + \int_\tau \bar a(\tau) (\partial_\tau + \Omega) a(\tau) \\
    & + g \int_\tau (a(\tau) + \bar a(\tau)) C_{x}(\tau) + \zeta \frac{g^2}{\Omega} \int_\tau C_{x}^2(\tau) \,.
\end{split}
\end{equation}
Here $\oint_{a, \bar a, c} \equiv \int \mathcal D[a, \bar a] \mathcal D[c]$ is the functional integral over coherent-state paths, with $a$ and $\bar a$ the complex conjugate fields corresponding to the cavity and $c$ standing for an unspecified matter coherent field, and $\int_\tau \equiv \int_0^\beta d\tau$ is the integral over imaginary time. A partial integration over the photonic fields, which is but a collection of Gaussian integrals, yields an effective action $S_{\rm eff} =  S_{\rm m} + S_{\rm ind}$ with an induced retarded collective interaction \cite{feynman1963the, grabert1988quantum} 
\begin{equation}
    S_{\rm ind} =  g^2 \int_{\tau, \tau'} C_{x}(\tau) D_0(\tau - \tau') C_{x}(\tau') + \zeta \frac{g^2}{\Omega} \int_\tau C_{x}^2(\tau)  \,.
    \label{eq:Seff}
\end{equation}
Here $D_0$ is the free photon propagator
\begin{equation}
    D_0(\tau) = -\frac{e^{-\tau \Omega}}{1 - e^{-\beta \Omega}} \,,
\end{equation}
or in Matsubara frequency space
\begin{equation}
    D_0(\omega_m) = \frac{1}{i \omega_m - \Omega} \,.
\end{equation}
It is convenient to write
\begin{equation}
    S_{\rm ind} = \frac{1}{2} \int_{\tau, \tau'} C_{x}(\tau) \frac{1}{N} V_{\rm ind}(\tau - \tau') C_{x}(\tau') \,,
    \label{eq:Sind}
\end{equation}
with 
\begin{equation}
    V_{\rm ind} (\tau) = \lambda^2 (D_0 (\tau) + D_0(\beta - \tau)) + 2 \zeta \frac{\lambda^2}{\Omega} \delta(\tau) \,,
\end{equation}
or in Matsubara frequency space
\begin{equation}
    V_{\rm ind} (\omega_m) = 2 \lambda^2 \frac{\Omega^2 (\zeta - 1) + \zeta \omega_m^2}{\Omega (\omega_m^2 + \Omega^2)} \,.
\end{equation}
Note that we symmetrize $V_{\rm ind}$ with respect to $\tau$. It can be seen that the odd part of $V_{\rm ind}$ does not contribute to the action. This will be justified later, when we relate $V_{\rm ind}$ to the  propagator of an auxiliary real scalar field, which must be even. 

\subsubsection{Auxiliary field decoupling}
The induced interaction can be decoupled with a Hubbard-Stratonovich transformation that introduces an auxiliary scalar field, $\varphi$:
\begin{equation}
\begin{split}
    e^{-S_{\mathrm{ind}}}=\frac{1}{Z_\varphi} \oint_{\varphi} \exp \Bigl(& -\frac{1}{2} N \int_{\tau, \tau'}  \varphi(\tau) V_{\mathrm{ind}}^{-1}(\tau-\tau^{\prime}) \varphi(\tau^{\prime}) \\
    & -i \int_\tau  \varphi(\tau) C_{x}(\tau) \Bigr) \,.
\end{split}
    \label{newHS}
\end{equation}
We define the propagator of the auxiliary field as
\begin{equation}
    W(\tau) = \langle \varphi(\tau) \varphi(0) \rangle^{\rm c} \,.
\end{equation}
The generating functional for bare connected correlation functions of the matter coupling operator, $C_{x}$, is $\mathcal G_{\rm m}^0[\xi] = - N^{-1} \log Z_{\rm m}[\xi]$ with
\begin{equation}
    Z_{\rm m}[\xi] = \oint_{c} e^{-(S_{\rm m} + i \int_\tau \xi(\tau) C_{x}(\tau))} \,, 
    \label{eq:generatingfuncfreematter}
\end{equation}
such that
\begin{equation}
\begin{split}
    \frac{\delta}{\delta \xi(\tau_1)} \cdots & \frac{\delta}{\delta \xi(\tau_n)} \mathcal G_{\rm m}^0[\xi] \Bigr|_{\xi = 0} = \\
    & = - \frac{(-i)^n}{N}  \langle C_{x}(\tau_1) \cdots C_{x}(\tau_n)\rangle^{\rm c}_{\rm m} \\
    & \equiv \chi_{xx, 0}^{(n)}(\tau_1, \ldots, \tau_n) \,. 
\end{split}
\end{equation}
In the following we will denote $\chi_{xx, 0}^{(2)} \equiv \chi_{xx, 0}$. With this, after partial integration over the matter degrees of freedom, we can write \cite{lenk2022collective}
\begin{equation}
    Z = \oint_{\varphi} e^{-N f[\varphi]}  \,,
\end{equation}
with
\begin{equation}
    f[\varphi] =  \frac{1}{2} \int_{\tau, \tau'} \varphi(\tau) V_{\mathrm{ind}}^{-1}(\tau-\tau^{\prime}) \varphi(\tau^{\prime} ) + \mathcal G_{\rm m}^0[\varphi] \,.
    \label{eq:f}
\end{equation}
The exponent is proportional to $N$, suggesting that it can be treated with the saddle-point method for large $N$. The number of terms required for a good approximation will be determined by the value of $N$. In the thermodynamic limit, $N \to \infty$, it will be justified to truncate the expansion at the quadratic term, which will reveal what is the free propagator of the auxiliary field $\varphi$. The condition that the functional derivative of $f$ with respect to $\varphi$ vanishes
\begin{equation}
\begin{split}
     \int_\tau^{\prime} V_{\rm ind}^{-1}(\tau-\tau^{\prime}) \varphi_{\rm sp}(\tau^{\prime}) + \frac{\delta \mathcal G_{\rm m}^0 [\varphi_{\rm sp}]}{\delta \varphi_{\rm sp}(\tau)} &=\\
     \int_\tau^{\prime} V_{\rm ind}^{-1}(\tau-\tau^{\prime}) \varphi_{\rm sp}(\tau^{\prime}) + \frac{i}{N} \langle C_{x} (\tau) \rangle_{\varphi_{\rm sp}} &= 0
\end{split}
\end{equation}
or in Matsubara frequency space
\begin{equation}
    \varphi_{\rm sp}(\omega_m) =  -\frac{i}{N} V_{\rm ind}(\omega_m) \langle C_{x} (\omega_m) \rangle_{\varphi_{\rm sp}}
    \label{eq:saddlepointcondition}
\end{equation}
formally defines $\varphi_{\rm sp}$. Let us postpone the issue of finding $\varphi_{\rm sp}$ for now and continue on by expanding $f$ in powers of $\varphi$ around $\varphi_{\rm sp}$. It is convenient to note that, by definition,
\begin{equation}
    \mathcal G_{\rm m}^0[\xi]=\sum_{n_{\rm{even}}} \frac{1}{n !} \int_{\tau_1, \ldots, \tau_n}  \xi\left(\tau_1\right) \cdots \xi\left(\tau_n\right) \chi_{xx,0}^{(n)}\left(\tau_1, \ldots, \tau_n\right) \,.
\end{equation}
By virtue of $\varphi_{\rm sp}$ being a minimum of $f$, $f$ is locally quadratic in its vicinity
\begin{equation}
\begin{split}
     f[\varphi] - f[\varphi_{\rm sp}] =& \frac{1}{2} \int_{\tau, \tau'} \delta \varphi(\tau) (N W_0)^{-1}(\tau - \tau') \delta \varphi(\tau') \\
     & + \mathcal O(\delta \varphi^4) \,.
\end{split}
\end{equation}
Here $\delta \varphi = \varphi - \varphi_{\rm sp}$ and
\begin{equation}
     (N W_0)^{-1} = V_{\rm ind}^{-1} + \tilde \chi_{xx,0}\,,
     \label{eq:W0identity}
\end{equation}
with 
\begin{equation}
\begin{split}
    \tilde \chi_{xx, 0} (\tau, \tau') = & \frac{\delta}{\delta \varphi_{\rm sp}(\tau)}\frac{\delta}{\delta \varphi_{\rm sp}(\tau')} \mathcal G_{\rm m}^0[\varphi_{\rm sp}] \\
     = & \frac{1}{N} \langle C_x(\tau) C_x(\tau')\rangle_{\varphi_{\rm{sp}}}^{\rm c} \,.
\end{split}
    \label{eq:chi0tilde}
\end{equation}
%
%
Here $\tilde \chi_{xx, 0}$ refers to the bare matter propagator but subject a field $i \varphi_{\rm sp}$, as per Eq. \eqref{eq:generatingfuncfreematter}. Note that if $\varphi_{\rm sp} = 0$ then $\tilde \chi_{xx, 0} = \chi_{xx, 0}$. At this point we note that $W_0$, which is the free propagator of $\varphi$, a real scalar field, is a function of $V_{\rm ind}$. It now clear why we symmetrize $V_{\rm ind}$, $W_0$ must be even in time. The importance of Eq. \eqref{eq:W0identity} stems from the fact that for $N \to \infty$ one can safely truncate the saddle-point expansion to second order, which implies that $W = W_0$. We can now obtain relations between the auxiliary field propagator, $W$, the photonic propagator, $D$, and matter response functions.

\subsubsection{Dressed response functions in terms of bare response functions}
In the following we will define generating functionals for connected correlators in imaginary time. These differ from the thermal Green function as defined in Eq. \eqref{eq:thermalgreen} in symmetry broken phases when the expectation value of the operator for which one is computing the connected correlator acquires a finite value. However, this difference only affects the zero frequency component of the correlator and is thus unimportant for the analytic continuation to real frequencies.

Let us define the generating functional for photonic connected correlators by introducing a complex field in Eq. \eqref{eq:ogaction}: $\mathcal G_{\rm ph} [\eta, \bar \eta] = - \log Z[\eta, \bar \eta]$, with
\begin{equation}
    Z[\eta, \bar \eta] = \oint_{a, \bar a, c} e^{-(S + \int_\tau (a(\tau) \bar \eta(\tau) + \bar a(\tau) \eta(\tau))} \,.
\end{equation}
After partial integration of the cavity fields, this yields
\begin{equation}
\begin{split}
    Z [\eta, \bar \eta] = \oint_{c} \exp - \Bigl( & S_{\rm m} + \zeta \frac{g^2}{\Omega} \int_\tau C_{x}^2(\tau) \\
    & + \int_{\tau, \tau'} \bar m(\tau) D_0(\tau - \tau') m(\tau) \Bigr) \,,
\end{split}
\end{equation}
with $m(\tau) = \eta(\tau) + g C_{x}(\tau)$. With this
\begin{equation}
\begin{split}
    D(\tau - \tau') =&  \left. \frac{\delta}{\delta \bar \eta(\tau)} \frac{\delta}{\delta \eta(\tau')} \mathcal G_{\rm ph} [\eta, \bar \eta] \right|_{\eta = \bar \eta = 0} \\
    =& D_0(\tau - \tau') \\
    & - \lambda^2 \int_{u, v} D_0(\tau - u) \chi_{xx}(u - v) D_0(v - \tau') \,,
\end{split}
\end{equation}
or in Matsubara frequency space
\begin{equation}
    D(\omega_m) = D_0(\omega_m) - \lambda^2 D_0(\omega_m) \chi_{xx}(\omega_m) D_0(\omega_m) \,.
    \label{eq:relpropphtonmatter}
\end{equation}
Likewise, we can define the generating functional for matter connected correlators. Since we are interested in response functions, i.e. two point-correlators, we can rather generally consider correlations of three distinct matter operators, the matter coupling operator, $C_x$, and two others, that we label $C_y$ and $C_z$. Note that $x, y$ and $z$ are simply indices here and they do not necessarily indicate spatial direction. The generating functional is $\mathcal G_{\rm m} [\xi_x, \xi_y, \xi_z] = - N^{-1} \log Z[\xi_x, \xi_y, \xi_z]$, with
\begin{equation}
    Z[\xi_x, \xi_y, \xi_z] = \oint_{c, \varphi} e^{-(S_{\rm eff} +i \int_\tau  \sum_{r} \xi_r(\tau) C_{r}(\tau))} \,.
\end{equation}
After partial integration over the matter fields, this yields
\begin{equation}
    Z[\xi_x, \xi_y, \xi_z] =\oint_{\varphi} e^{-N f[\varphi, \xi_x, \xi_y, \xi_z]}\,,
    \label{eq:generatingfuncmatter2}
\end{equation}
with
\begin{equation}
\begin{split}
    f[\varphi, \xi_x, \xi_y, \xi_z] = &  \frac{1}{2} \int_{\tau, \tau'} \varphi(\tau) V_{\mathrm{ind}}^{-1}(\tau-\tau^{\prime}) \varphi(\tau^{\prime} ) \\
    & + \mathcal G_{\rm m}^0[\xi_x + \varphi, \xi_y, \xi_z] \,,
\end{split}
\end{equation}
Here $\mathcal G_{\rm m}^0 [\xi_x, \xi_y, \xi_z] = - N^{-1} \log Z_{\rm m}[\xi_x, \xi_y, \xi_z]$ is the generalization of $\mathcal G_{\rm m}^0[\xi]$ \eqref{eq:generatingfuncfreematter}, with
\begin{equation}
   Z_{\rm m} [\xi_x, \xi_y, \xi_z] = \oint_{c} e^{-(S_{\rm m} +i \int_\tau  \sum_{r} \xi_r(\tau) C_{r}(\tau))} \,.
    \label{eq:extendedgeneratingfuncfreematter}
\end{equation}
Then a second-order expansion of $f[\varphi, \xi_x, \xi_y, \xi_z]$ around $\varphi_{\rm sp}$ yields
\begin{widetext}
    \begin{equation}
\begin{split}
    f[\varphi, \xi_x, \xi_y, \xi_z] - f[\varphi_{\rm sp}, 0, 0, 0]  =  \frac{1}{2} \int_{\tau, \tau'} \Bigl( &\delta \varphi(\tau) (NW)^{-1}(\tau - \tau')\delta \varphi(\tau') + \sum_{r s} \xi_{r}(\tau) \tilde \chi_{r s, 0}(\tau - \tau') \xi_{s}(\tau') \\
    & + \, \delta\varphi(\tau) \sum_r \tilde \chi_{x r,0}(\tau - \tau') \xi_{r} (\tau') + \sum_r \xi_{r} (\tau) \tilde \chi_{r x, 0}(\tau - \tau') \delta \varphi(\tau) \Bigr) \,.
\end{split}
\end{equation}
The functional integral over the auxiliary-field displacements $\delta \varphi$ is just a Gaussian integral that we can perform, yielding 
\begin{equation}
\begin{split}
    \mathcal G_{\rm m} [\xi_x, \xi_y, \xi_z] = {\rm cst.} + \frac{1}{2} \sum_{r s} \int_{\tau, \tau'} \xi_{r}(\tau)\Biggl(& \tilde \chi_{r s, 0}(\tau - \tau') - \int_{u, v} \tilde \chi_{r x, 0}(\tau - u) N W(u - v) \tilde \chi_{x s, 0}(u - \tau') \Biggr) \xi_{s}(\tau')\,.
\end{split}
\end{equation}
With this
\begin{equation}
\begin{split}
    \chi_{r s} (\tau - \tau') = \left. \frac{\delta}{\delta \xi_{r}(\tau)} \frac{\delta}{\delta \xi_{s}(\tau')} \mathcal G_{\rm m} [\xi_x, \xi_y, \xi_z] \right|_{\xi_r = 0 \; \forall r} 
     = \tilde \chi_{r s, 0}(\tau - \tau') - \int_{u, v} \tilde \chi_{r x, 0}(\tau - u) N W(u - v) \tilde \chi_{x s, 0}(u - \tau') \,,
\end{split}
\end{equation}
\end{widetext}
or in Matsubara frequency space
\begin{equation}
    \chi_{r s} (\omega_m) = \tilde \chi_{r s, 0} (\omega_m) - \tilde \chi_{r x, 0}(\omega_m) N W (\omega_m) \tilde \chi_{x r,0}(\omega_m) \,.
    \label{eq:relpropmatteraux}
\end{equation}
Putting together Eqs. \eqref{eq:W0identity} and \eqref{eq:relpropmatteraux} we arrive at the relations
\begin{align}
    & \chi_{xr}(\omega_m) = \frac{\tilde \chi_{xr,0}(\omega_m)}{1 + V_{\rm ind}(\omega_m) \tilde \chi_{xx, 0}(\omega_m)} \label{eq:chixaidentity} \,, \\
    & \chi_{r x}(\omega_m) = \frac{\tilde \chi_{r x, 0}(\omega_m)}{1 + V_{\rm ind}(\omega_m) \tilde \chi_{xx, 0}(\omega_m)} \label{eq:chiaxidentity} \,,\\
    \begin{split}
        & \left. \chi_{r s} \right|_{r, s \neq x}(\omega_m) = \tilde \chi_{r s, 0}(\omega_m) \\
        & \qquad \qquad \qquad \quad \; - \frac{\tilde \chi_{r x, 0}(\omega_m) V_{\rm ind}(\omega_m) \tilde \chi_{x s, 0}(\omega_m)}{1 + V_{\rm ind}(\omega_m) \tilde \chi_{xx, 0}(\omega_m)} \,,
    \end{split}
     \label{eq:chiyyidentity}
\end{align}
which together with Eq. \eqref{eq:relpropphtonmatter} give us the dependence of the photonic propagator and matter response functions on the bare matter response function and the free photon propagator. Note from Eqs. \eqref{eq:chi0tilde} and \eqref{eq:extendedgeneratingfuncfreematter} that $\tilde \chi_{r s, 0}$ is simply the response function of the bare matter subject to an external field $\varphi_{\rm sp}$, the saddle point. 
Equations \eqref{eq:chixaidentity}, \eqref{eq:chiaxidentity} and \eqref{eq:chiyyidentity} exhibit a structure similar to the response functions commonly derived using the random phase approximation, such as those found in \cite{lenk2022collective}. It is important to highlight that these equations remain valid for all parameter ranges and are applicable to both normal and superradiant phases.

\subsubsection{Computing the saddle point $\varphi_{\rm sp}$.}
We look for constant solutions of the saddle-point equations, i.e. $\varphi_{\rm sp}(\tau) = \varphi_{\rm sp} = \varphi_{\rm sp}(\omega_m = 0)$ (all other frequency components being zero).
Then, from Eq. \ref{eq:saddlepointcondition} we find
\begin{equation}
    \varphi_{\rm sp} = -\frac{i}{N} V_{\rm ind}(\omega_m = 0) \langle C_x \rangle_{\varphi_{\rm sp}} \,.
    \label{eq:constphi0condition}
\end{equation}
This tells us that $\varphi_{\rm sp}$ is self-consistently proportional to $\langle C_x \rangle_{\varphi_{\rm sp}}$, i.e. to the expectation value of $C_x$ for the bare matter subject to a field $i \varphi_{\rm{sp}} =\frac{1}{N} V_{\rm ind}(\omega_m = 0) \langle C_x \rangle$. This is precisely the self-consistent condition that arises from computing $\langle C_x\rangle$ from the mean-field Hamiltonian 
\begin{equation}
    H_{\rm eff}^{\rm MF} = H_{\rm m} + \frac{2\lambda^2 (\zeta - 1)}{N \Omega} \langle C_x \rangle C_x - \frac{\lambda^2 (\zeta - 1)}{N \Omega} \langle C_x \rangle^2 \,.
    \label{eq:Heff}
\end{equation}
This is the mean-field theory of the effective Hamiltonian that arises from taking the static limit in $V_{\rm ind}$ in the effective action \cite{romanroche2022effective}.
Thus, if we assume $\varphi_{\rm sp}$ constant, $\tilde \chi_{rs, 0}$ are just the response functions of the mean field matter Hamiltonian of Eq. \eqref{eq:Heff}. Thus, $\tilde \chi_{rs, 0}$ are obtained as functions of $\langle C_x \rangle$ which can be computed by solving $H_{\rm eff}^{\rm MF}$ variationaly with respect to $\langle C_x \rangle$. Note that in the absence of symmetry breaking, i.e. when $\langle C_x \rangle =0$, or if $\zeta = 1$, the effective matter Hamiltonian is just the bare matter Hamiltonian and thus $\tilde \chi_{r s, 0} = \chi_{r s, 0}$ is just the bare matter response.

\subsection{Equations of motion for the retarded Green functions}

An alternative way to compute retarded Green functions is by formulating their equations of motion and, under some assumptions, solving them. Defining $G^{\rm r}_{A, B}(t) \equiv G^{\rm r}_{A, B}(t, 0)$ we have from Eq. \eqref{eq:retardedgreenfunction} that
\begin{equation}
    i \partial_t G^{\rm r}_{A, B}(t) = \delta(t) \langle [A, B] \rangle - G^{\rm r}_{[H, A], B}(t)
\end{equation}
or in frequency space
\begin{equation}
    \omega_+ G^{\rm r}_{A, B}(\omega) = \langle [A, B] \rangle - G^{\rm r}_{[H, A], B}(\omega) \,,
\end{equation}
with $\omega_+ = \omega + i0^+$.

\subsubsection{Solving the equations of motion}
Let $\{|\alpha\rangle\}$ be a basis of the Hilbert space of the matter subsystem and let us define the Hubbard operators $X^{\alpha \gamma} = |\alpha \rangle \langle \gamma|$. We start by computing the matter correlation function $\chi_{rs}$. Note that 
\begin{equation}
    \chi_{rs} = -\frac{1}{N} G^{\rm r}_{C_r, C_s} = -\frac{1}{N} \sum_{\alpha, \gamma} C_r^{\alpha \gamma} G^{\rm r}_{X^{\alpha \gamma}, C_s}
    \label{eq:mattergreenfunctiondecomp}
\end{equation}
where $C_r^{\alpha \gamma} = \langle \alpha | C_r | \gamma \rangle$. Then
\begin{equation}
\begin{split}
     \omega_+ G^{\rm r}_{X^{\alpha \gamma}, C_s}(\omega) =& \langle [X^{\alpha \gamma }, C_s] \rangle - G^{\rm r}_{[H, X^{\alpha \gamma}], C_s} (\omega) \\
     =& -\Gamma_s^{\alpha \gamma} - G^{\rm r}_{[H_{\rm m}, X^{\alpha \gamma}], C_s}(\omega) \\
     &- g G^{\rm r}_{(a + a^\dagger) [C_x, X^{\alpha \gamma}], C_s}(\omega) \\
     & - \zeta \frac{g^2}{\Omega} G^{\rm r}_{[C_x^2, X^{\alpha \gamma}], C_s}(\omega) \,.
     \label{eq:eommattergreenfucntion}
\end{split}
\end{equation}
Here $\Gamma_s^{\alpha \gamma} = \langle [C_s, X^{\alpha \gamma }] \rangle = \sum_\mu (C_s^{\mu \alpha} \langle X^{\mu \gamma} \rangle - C_s^{\gamma \mu} \langle X^{\alpha \mu} \rangle)$. The last two terms of Eq. \eqref{eq:eommattergreenfucntion} correspond to three-point Green functions. One could now write the equations of motion for such three-point Green functions, which would in turn depend on four-point Green functions, and so on and so forth. This infinite hierarchy of $n$-point correlators involves light-matter interactions, as discussed in the context of lasing physics \cite[Chapter 9]{gardiner2004quantum} or light emission \cite{delvalle2009Luminescence}. At some point, the hierarchy needs to be truncated. The crudest approximation is to restrict the theory to two-point Green functions, assuming that $G^{\rm r}_{AB, C} \approx \langle A \rangle G^{\rm r}_{B, C} +  \langle B \rangle G^{\rm r}_{A, C}$, which amounts to neglecting correlations of order higher than two for all terms stemming from light-matter interaction. In terms of fluctuations, this means that only linear fluctuations between the two systems are relevant, and that correlated fluctuations are largely suppressed.
This approach is reminiscent of RPA or mean-field approximations, which need to be justified. This has indeed been discussed within the Dicke model, where the traditional educated guess that quantum fluctuations vanish in the \(N \to \infty\) limit \cite[Chapter 9]{gardiner2004quantum} [Cf. with the equilibrium argument in \cite{wang1973phase}] has been rigorously demonstrated recently \cite{carollo2021exactness}. In that work, the proof was restricted to the Dicke model and within the Lindblad framework, \emph{i.e.}, adding dissipation in a Markovian manner. This result is relevant to us since LRT falls within the conditions of the theorem demonstrated in that reference, which requires the dynamics to start with an uncorrelated initial condition between light and matter. This can be safely assumed here since the response is independent of the excitation and can be obtained from excitation functions where the initial condition is assumed to be at equilibrium, which can be well approximated by uncorrelated light-matter states \cite{andolina2019cavity, romanroche2021photon}.
It is true, however, that in our case we are not considering dissipative Lindblad dynamics and we include potential matter-matter interactions. In any case, we believe that restricting to two-point Green functions can be proven to be exact, as it agrees with the saddle-point solution for $N \to \infty$  discussed in the previous section. In fact, this section concludes by obtaining the same response functions as within the path integral formalism.  This suggests that generalizing Ref. \cite{carollo2021exactness} might be possible in a future work.
With this,
\begin{equation}
    \begin{split}
        \omega_+ G^{\rm r}_{X^{\alpha \gamma}, C_s}(\omega) = & -\Gamma_s^{\alpha \gamma} - G^{\rm r}_{[H_{\rm m}, X^{\alpha \gamma}], C_s}(\omega) \\
        & - g \langle a + a^\dagger \rangle G^{\rm r}_{[C_x, X^{\alpha \gamma}], C_s}(\omega) \\
        & - \zeta \frac{2 g^2}{\Omega} \langle C_x \rangle G^{\rm r}_{[C_x, X^{\alpha \gamma}], C_s}(\omega) \\
        &- g \Gamma_x^{\alpha \gamma}\left(G^{\rm r}_{a, C_s}(\omega) + G^{\rm r}_{a^\dagger, C_s}(\omega)\right) \\
        & - \zeta \frac{2 g^2}{\Omega} \Gamma_x^{\alpha \gamma} G^{\rm r}_{C_x, C_s}(\omega) \,.
    \end{split}
\end{equation}
We can put together the second, third and fourth elements in the right-hand side to build a mean field matter Hamiltonian
\begin{equation}
    H_{\rm m}^{\rm MF} = H_{\rm m} + g \langle a + a^\dagger\rangle C_x + \zeta \frac{2 g^2}{\Omega} \langle C_x \rangle C_x \,.
\end{equation}
Note that the constant terms in this mean-field Hamiltonian are omitted here. We will show later that this Hamiltonian is equivalent to the mean-field effective Hamiltonian \eqref{eq:Heff} obtained with the path integral approach under a mean-field decoupling of full light-matter Hamiltonian. At this point we can assume that the basis that we have been using, $\{|\alpha\rangle\}$, is precisely the eigenbasis of $H_{\rm m}^{\rm MF}$, with eigenenergies $\{E_\alpha\}$, obtaining
\begin{equation}
    \begin{split}
        \omega_+ G^{\rm r}_{X^{\alpha \gamma}, C_s}(\omega) =&  -\Gamma_s^{\alpha \gamma} - (E_\alpha - E_\gamma) G^{\rm r}_{X^{\alpha \gamma}, C_s}(\omega) \\
        & - g \Gamma_x^{\alpha \gamma}\left(G^{\rm r}_{a, C_s}(\omega) + G^{\rm r}_{a^\dagger, C_s}(\omega)\right) \\
        & - \zeta \frac{2 g^2}{\Omega} \Gamma_x^{\alpha \gamma} G^{\rm r}_{C_x, C_s}(\omega) \,.
    \end{split}
    \label{eq:Xprop}
\end{equation}
We can compute the photon-matter Green functions that appear on the second line obtaining
\begin{align}
    (\omega_+ - \Omega) G^{\rm r}_{a, C_s}(\omega) &= g G^{\rm r}_{C_x, C_s}(\omega) \,, \label{eq:aprop}\\
    (\omega_+ + \Omega) G^{\rm r}_{a^\dagger, C_s}(\omega) &= -g G^{\rm r}_{C_x, C_s}(\omega) \,. \label{eq:adaggerprop}
\end{align}
Putting together Eqs. \eqref{eq:Xprop}, \eqref{eq:aprop} and \eqref{eq:adaggerprop} we get
\begin{widetext}
    \begin{equation}
    (\omega_+ + E_\alpha - E_\gamma) G^{\rm r}_{X^{\alpha \gamma}, C_s}(\omega) = -\Gamma_s^{\alpha \gamma} - \Gamma_x^{\alpha \gamma} V_{\rm ind}(\omega) \frac{1}{N} G^{\rm r}_{C_x, C_s}(\omega) \,,
\end{equation}
with $V_{\rm ind} (\omega) = V_{\rm ind} (\omega_m) |_{i \omega_m \to \omega + i0^+}$. Together with Eq. \eqref{eq:mattergreenfunctiondecomp} this gives
\begin{equation}
\begin{split}
        \chi_{rs}(\omega) =  \frac{1}{N} \sum_{\alpha \gamma} C_r^{\alpha \gamma} \frac{\Gamma_s^{\alpha \gamma}}{\omega_+ + E_\alpha - E_\gamma} - \frac{1}{N} \sum_{\alpha \gamma} C_r^{\alpha \gamma} \frac{\Gamma_x^{\alpha \gamma}}{\omega_+ + E_\alpha - E_\gamma} V_{\rm ind}(\omega) \chi_{x s}(\omega) \,.
\end{split}
    \label{eq:secondtolasteqforchi}
\end{equation}
From the definition of $\Gamma_s^{\alpha \gamma}$ and noting that $\langle X^{\alpha \gamma} \rangle = Z^{-1} \delta_{\alpha \gamma} \exp(-\beta E_\alpha)$ we find that
\begin{equation}
    \frac{1}{N} \sum_{\alpha \gamma} C_r^{\alpha \gamma} \frac{\Gamma_s^{\alpha \gamma}}{\omega_+ + E_\alpha - E_\gamma} = \frac{1}{N}\frac{1}{Z}  \sum_{\alpha \gamma} \frac{e^{-\beta E_\gamma} - e^{-\beta E_\alpha}}{\omega_+ + E_\alpha - E_\gamma} \langle \alpha | C_r | \gamma \rangle \langle \gamma | C_s | \alpha \rangle = \tilde \chi_{rs, 0}(\omega) \,.
\end{equation}
\end{widetext}
With this and Eq. \eqref{eq:secondtolasteqforchi}, we finally obtain 
    \begin{align}
    & \chi_{xr}(\omega) = \frac{\tilde \chi_{xr,0}(\omega)}{1 + V_{\rm ind}(\omega) \tilde \chi_{xx, 0}(\omega)} \,, \\
    & \chi_{r x}(\omega) = \frac{\tilde \chi_{r x, 0}(\omega)}{1 + V_{\rm ind}(\omega) \tilde \chi_{xx, 0}(\omega)} \,, \\
\begin{split}
    & \left. \chi_{rs} \right|_{r, s \neq x}(\omega) = \tilde \chi_{r s, 0}(\omega) \\
    & \qquad \qquad \qquad \; \; \; - \frac{\tilde \chi_{r x, 0}(\omega) V_{\rm ind}(\omega) \tilde \chi_{x s, 0}(\omega)}{1 + V_{\rm ind}(\omega) \tilde \chi_{xx, 0}(\omega)} \,, 
\end{split}
\end{align}
which coincide with the analytic continuation of Eqs. \eqref{eq:chixaidentity}, \eqref{eq:chiaxidentity} and \eqref{eq:chiyyidentity}.

Proceeding in a similar fashion, we can now compute $D$ and $D_+ = G_{a^\dagger, a^\dagger}^{\rm r}$. First we have
\begin{align}
    (\omega_+ - \Omega) D(\omega) &= 1 + g G^{\rm r}_{C_x, a^\dagger}(\omega) \,, \label{eq:D}\\
    (\omega_+ + \Omega) D_+(\omega) &= -g G^{\rm r}_{C_x, a^\dagger}(\omega) \,. \label{eq:Dplus}
\end{align}
Again we use the decomposition $G^{\rm r}_{C_x, a^\dagger} = \sum_{\alpha \gamma} C_x^{\alpha \gamma} G^{\rm r}_{X^{\alpha \gamma}, a^\dagger}$ and compute
\begin{equation}
\begin{split}
    (\omega_+ + E_\alpha - E_\gamma) G^{\rm r}_{X^{\alpha \gamma}, a^\dagger} = & -g \Gamma_x^{\alpha \gamma} \left(D(\omega) + D_+(\omega)\right) \\
    & - \zeta \frac{2 g^2}{\Omega} \Gamma_x^{\alpha \gamma} G^{\rm r}_{C_x, a^\dagger}(\omega) \,,
\end{split}
\end{equation}
such that
\begin{equation}
    G^{\rm r}_{C_x, a^\dagger} = \frac{-N g \tilde \chi_{xx, 0}(\omega)}{1 + \zeta \frac{2 \lambda^2}{\Omega} \tilde \chi_{xx, 0}(\omega)} \left(D(\omega) + D_+(\omega)\right) \,.
\end{equation}
Plugging this into Eqs. \eqref{eq:D} and \eqref{eq:Dplus} we get
\begin{align}
\begin{split}
    (\omega - \Omega) D(\omega) & = 1  \\
    & \quad - \frac{\lambda^2 \tilde \chi_{xx, 0}(\omega)}{1 + \zeta \frac{2 \lambda^2}{\Omega} \tilde \chi_{xx, 0}(\omega)} \left(D(\omega) + D_+(\omega)\right) \,,
\end{split}
\\
    (\omega + \Omega) D_+(\omega) &= \frac{\lambda^2 \tilde \chi_{xx, 0}(\omega)}{1 + \zeta \frac{2 \lambda^2}{\Omega} \tilde \chi_{xx, 0}(\omega)} \left(D(\omega) + D_+(\omega)\right) \,. 
\end{align}
We can now solve for $D$ and $D_+$ to get
\begin{align}
    D(\omega) &= \frac{\omega_+ + \Omega + \lambda^2 \frac{2 \zeta( \omega_+ + \Omega) - \Omega}{\Omega} \tilde \chi_{xx, 0}(\omega)}{\omega_+^2 - \Omega^2 + 2\lambda^2 \frac{\zeta(\omega_+^2 - \Omega^2) + \Omega^2}{\Omega} \tilde \chi_{xx, 0}(\omega)} \,, \label{eq:Dfromeom}\\
    D_+(\omega) &= \frac{\lambda^2 \tilde \chi_{xx, 0}(\omega)}{\omega_+^2 - \Omega^2 + 2\lambda^2 \frac{\zeta(\omega_+^2 - \Omega^2) + \Omega^2}{\Omega} \tilde \chi_{xx, 0}(\omega)}\,. \label{eq:Dpfromeom}
\end{align}
Although not immediately obvious, Eq. \eqref{eq:Dfromeom} is precisely the analytic continuation of Eq. \eqref{eq:relpropphtonmatter}. We prove this relation in App. \ref{app:equivalence}.

\subsubsection{Mean-field decoupling of the light-matter Hamiltonian}

Let us consider the Hamiltonian in Eq. \eqref{eq:Hstart} with a mean-field decoupling of the light-matter interaction and $P^2$ terms. We expand $C_x = \langle C_x\rangle + \delta C_x$ and $a + a^\dagger = \langle a + a^\dagger \rangle + \delta(a + a^\dagger)$ and neglect terms quadratic in fluctuations, obtaining two decoupled Hamiltonians and some constants: $H = H_{\rm m}^{\rm{MF}} + H_{\rm{ph}}^{\rm{MF}} + \rm{cst.}$ with
\begin{align}
    & H_{\rm m}^{\rm{MF}} = H_{\rm m} + g \langle a + a^\dagger \rangle C_x + \zeta \frac{2g^2}{\Omega} \langle C_x\rangle C_x   \,, \\
    & H_{\rm{ph}}^{\rm{MF}} = \Omega a^\dagger a + g(a + a^\dagger) \langle C_x \rangle \,, \\
    & {\rm cst}. = - g \langle a + a^\dagger \rangle \langle C_x \rangle - \zeta \frac{g^2}{\Omega} \langle C_x\rangle^2  \,.
\end{align}
The mean-field photonic Hamiltonian $H_{\rm{ph}}^{\rm{MF}}$ can be diagonalized with a displacement, $a = b - g / \Omega \langle C_x \rangle$, yielding $H_{\rm{ph}}^{\rm{MF}} = \Omega b^\dagger b$, a new constant $- g^2/\Omega \langle C_x\rangle^2$ and the relation $\langle a + a^\dagger \rangle = - 2g / \Omega \langle C_x \rangle$. Plucking this relation into the mean-field matter Hamiltonian, $H_{\rm m}^{\rm{MF}}$, and all the constants gathered in the different steps yields precisely the mean-field effective Hamiltonian, $H_{\rm eff}^{\rm MF}$, of Eq. \eqref{eq:Heff}. The constants are important for a potential variational solution.

\subsection{Comparing the two theories}

In this section we have shown the equivalence between two different approaches to address the linear response regime of cavity QED materials. Although seemingly very different, the path-integral in imaginary time and the equation of motion in real time provide exact analytical results in the thermodynamic limit, $N\to\infty$. The effect of the thermodynamic limit is clearly identified in the path integral approach when the saddle point method is invoked. Instead, with the equation of motion method it requires an analysis of the scaling of the correlated parts of higher-order correlation functions with $N$.
On the other hand, the equations-of-motion approach does not require the auxiliary field of the Hubbard-Stratonovich transformation, which allows us to interpret the steps used to calculate its propagator $W$. We can see that the auxiliary field $\varphi_{\text{sp}}$ plays the same role as the mean-field background in the equation of motion, which couples to the fluctuations.

The two approaches cement the understanding that mean-field approaches are exact in the thermodynamic limit for Dicke-like models, even if matter interactions are considered.
Together, they paint the following picture.
One can construct a hierarchy of light-matter correlations: observables, response functions, three-point correlators, etc.. Normally, this hierarchy is not closed, meaning that $n$-point correlators depend on $n+1$-point correlators, and so on, ad infinitum.
This can be seen in our equations-of-motion approach.
In contrast, for Dicke-like models of the form of Eq. \eqref{eq:Hstart} it seems that this hierarchy can be truncated at any order and still provide exact results for that order of the hierarchy.
In this paper we have confirmed that this is the case for one- and two-point correlators.
One-point correlators (expected values of a single light or matter observable) are computed from the effective mean-field Hamiltonian of Eq. \eqref{eq:Heff}.
Note that at this level it seems like we are neglecting all light-matter correlations, even two-point correlations.
Then, two-point correlations, which define linear-response theory, are given by Eqs. \eqref{eq:relpropphtonmatter}, \eqref{eq:chixaidentity}, \eqref{eq:chiaxidentity} and \eqref{eq:chiyyidentity} and depend solely on bare two-point correlators, in particular on those of the mean-field effective Hamiltonian.
From our path-integral approach, we see that this is simply a consequence of the fact that each order of correlations corresponds to a new term of the saddle point expansion, and as such is a factor $1/N$ less significant than the previous order \cite{bacciconi2023firstorder}. 
In the thermodynamic limit every order is completely independent of all higher orders in the hierarchy.

As a caveat, note that this discussion only applies to light-matter correlations.
The material subsystem will generally be an interacting many-body system on its own where correlations can not be computed exactly with the same mean-field philosophy. 
As such, even though we have expressed the response functions of Dicke-like models in terms of the bare response functions of the cavity and the mean-field effective Hamiltonian, computing the latter will generally be a challenge itself. In the rest of the paper and in Ref. \cite{romanroche2024linear} we study some cases in which this can be done exactly as well.

This has been studied and exploited before at equilibrium \cite{wang1973phase, lee2004firstorder, andolina2019cavity, romanroche2021photon, romanroche2022effective}, in the dynamics \cite{carollo2021exactness,rincon2024mumax3cqed} and in Refs. \cite{lenk2022collective, lenk2022dynamical} and now here for the LRT. Beyond the LRT, the equations-of-motion approach opens the possibility of studying real-time dynamics. 

Finally, the comparison between the two approaches is not just an academic exercise. In many cases it can be interesting to go beyond the $N\to\infty$ limit. This would be the case in topological systems, where boundaries play an important role. In that case the two approaches offer complementary methods incorporate finite-size corrections. In the path integral approach, corrections are computed by extending the saddle-point expansion to higher orders, and thus relaxing $W = W_0$ for the weaker $W^{-1} = W_0^{-1} + \Pi$, where $\Pi$ can be computed with standard diagrammatic techniques in terms of fourth- and higher-order matter response functions \cite{lenk2022collective}. In the equations-of-motion approach, corrections are introduced by expanding the theory from two- to three- and higher-order Green functions before invoking the mean-field factoring of the correlators.

\section{Integer quantum Hall effect in a cavity}
\label{qhe}
\subsection{The Hamiltonian}
To demonstrate the applicability of our theory, we will use it to study the modification of the integer quantum Hall effect. The matter system under consideration is a two-dimensional electron gas subject to a perpendicular classical magnetic field. This model has been used to explain the breakdown of topological protection that underlies the observed modifications of the quantum Hall effect by coupling to a cavity \cite{ciuti2021cavitymediated, appugliese2022breakdown, rokaj2022polaritonic, rokaj2023weakened}. The Hamiltonian of the electron gas coupled to a single-mode cavity reads
\begin{equation}
\begin{split}
    H =&  \sum_j \frac{(\boldsymbol p_j - e \boldsymbol A_{\rm ext}(\boldsymbol r_j) - e \boldsymbol A_0 (b + b^\dagger))^2}{2m} \\
    & + \sum_{i > j} V(\boldsymbol r_i - \boldsymbol r_j) + \Omega b^\dagger b \,.
\end{split}
    \label{eq:Hqhe}
\end{equation}
Here $\boldsymbol r_i$, $\boldsymbol p_i$ are respectively the position and momentum operators of the $i$-th electron. Both the external classical vector potential and the cavity's vector potential point along the $x$-direction: $\boldsymbol A_{\rm ext} (\boldsymbol r_j) = -B y_j \boldsymbol e_x$ and $\boldsymbol A_0 = \sqrt{1/(2 \epsilon_0 V \Omega)} \boldsymbol e_x$. The $V$ and $\epsilon_0$ are respectively the cavity mode volume and the dielectric constant. $V(\boldsymbol r_i - \boldsymbol r_j)$ is the Coulomb interaction between the electrons. After some manipulations, we can write the Hamiltonian as 
\begin{equation}
\begin{split}
     H = & H_{\rm m} - \frac{\omega_{\rm p}}{\sqrt{N}} \left( \sqrt{\frac{\omega_{\rm p}}{\Omega}} \bar p_x + \sqrt{\frac{\omega_{\rm c}}{\Omega}} \bar y\right) \left(b + b^\dagger \right) \\
     & + \Omega b^\dagger b + \frac{\omega_{\rm p}^2}{4 \Omega} \left(b + b^\dagger \right)^2 \,,
\end{split}
\end{equation}
where $H_{\rm m}$ is the bare (without cavity) electron gas Hamiltonian
\begin{equation}
    H_{\rm m} = \sum_j \frac{(\boldsymbol p_j - e \boldsymbol A_{\rm ext}(\boldsymbol r_j))^2}{2m} + \sum_{i > j} V(\boldsymbol r_i - \boldsymbol r_j) 
\end{equation}
and $\bar p_x$ and $\bar y$ are adimensionalized collective operators
\begin{align}
    & \bar p_x = \frac{1}{\sqrt{2 m \omega_{\rm p}}} \sum_j p_{x, j} \,, \\
    & \bar y = \sqrt{\frac{m \omega_{\rm c}}{2}} \sum_j y_j \,.
\end{align}
We have defined the plasma frequency, $\omega_{\rm p} = \sqrt{e^2 \rho / (m \epsilon_0)}$ with $\rho = N / V$ the density of electrons is the cavity, and the cyclotron frequency $\omega_{\rm c} = e B / m$. The photonic terms of the Hamiltonian can be diagonalized with a Bogoliubov transformation, which yields
\begin{equation}
    H = H_{\rm m} - \frac{\omega_{\rm p}}{\sqrt{N}} \left( \sqrt{\frac{\omega_{\rm p}}{\tilde \omega}} \bar p_x + \sqrt{\frac{\omega_{\rm c}}{\tilde \omega}} \bar y\right) \left(a + a^\dagger \right) + \tilde \omega a^\dagger a \,,
    \label{eq:compactHqhe}
\end{equation}
with $\tilde \omega^2 = \Omega^2 + \omega_{\rm p}^2$.
By identifying $g = - \omega_{\rm p} / \sqrt{N}$ and 
$C_{x} = \sqrt{\omega_{\rm p} / \tilde \omega} \bar p_x + \sqrt{\omega_{\rm c} / \tilde \omega} \bar y$ the Hamiltonian takes the form of Eq.  \eqref{eq:Hstart}.
Note that in this case there is no $P^2$ term, i.e. $\zeta = 0$. In exchange, the elimination of the $A^2$ term through the Bogoliubov transform renormalizes the cavity frequency and light-matter coupling. 

\subsection{Current and optical conductivity}
In the quantum Hall effect \cite{vonklitzing1986the, vonklitzing202040}, one is typically interested in computing the longitudinal and transverse conductivities. The optical conductivity tensor can be computed with the Kubo formalism in terms of the current response functions
\begin{equation}
    \sigma_{rs}(\omega) = \frac{i}{\omega_+} \left[\frac{e^2 \rho_{\rm{2D}}}{m} \delta_{r s} + \frac{G^{\rm r}_{J_r, J_s}(\omega)}{A} \right] \,,
\end{equation}
where $\omega_+ = \omega + i \delta$, $A$ and $\rho_{\rm{2D}}$ are the area and surface density of the two-dimensional electron gas, $\delta$ is the broadening parameter, $\delta_{rs}$ is the Kronecker delta and $J_{r, s}$ are the current operators in the $r$ and $s$ directions, with $r, s \in \{x, y\}$ \cite{kubo1957statisticalmechanical, giuliani2005quantum}. We will now compute $G^{\rm r}_{J_r, J_s}$ using the theory outlined in Sec. \ref{LRT}, i.e. we will express $G^{\rm r}_{J_r, J_s}$ in terms of the response functions of the bare electron gas, $G^{r0}_{J_r, J_s}$.

The current operator is given by 
\begin{equation}
\begin{split}
     \boldsymbol{J} & = e \sum_j \boldsymbol v_j \\
     & = \frac{e}{m} \sum_j \left(\boldsymbol p_j - e \boldsymbol A_{\rm ext}(\boldsymbol r_j) - e \boldsymbol A_0 (b + b^\dagger) \right) \,.
\end{split}
    \label{eq:Jwithcavity}
\end{equation}
Then, after some manipulation, we can express
\begin{align}
    & J_x = \sqrt{\frac{2 e^2 \tilde \omega}{m}} \left( C_x + \gamma (a + a^\dagger)\right) \,, \label{eq:Jx}\\
    & J_y = \sqrt{\frac{2 e^2 \tilde \omega}{m}} C_y \,, \label{eq:Jy}
\end{align}
with $\gamma = -\sqrt{N} \omega_{\rm p} / (2 \tilde \omega)$ and $C_{y} = \sqrt{\omega_{\rm p} / \tilde \omega} \bar p_y$. Here $\sqrt{2m\omega_{\rm p}} \bar p_y = \sum_j p_{y, j}$. Meaning that we need to compute
\begin{align}
   & G^{\rm r}_{J_x, J_x} = \frac{2 e^2 \tilde \omega}{m}  G^{\rm r}_{C_x + \gamma (a + a^\dagger), C_x + \gamma (a + a^\dagger)} \,, \label{eq:mixed1}\\
   & G^{\rm r}_{J_x, J_y} = \frac{2 e^2 \tilde \omega}{m}  G^{\rm r}_{C_x + \gamma (a + a^\dagger), C_y} \,, \label{eq:mixed2} \\
   & G^{\rm r}_{J_y, J_y} = \frac{2 e^2 \tilde \omega}{m}  G^{\rm r}_{C_y, C_y} \,. \label{eq:mixed3}
\end{align}
Following App. \ref{app:currentcorrelatorsqhe}, these are given by
\begin{widetext}
\begin{align}
    & G^{\rm r}_{J_x, J_y} (\omega) = \left(1 + \left(\frac{\omega_{\rm p}}{\tilde \omega}\right)^2 \frac{\tilde \omega}{2} D_0^{\rm s} (\omega)\right) \frac{ G^{\rm r, 0}_{J_x, J_y}  (\omega)}{1 - V_{\rm ind}(\omega) \frac{m}{2 e^2 \tilde \omega N}  G^{\rm r, 0}_{J_x, J_x}(\omega)} \,,
    \label{eq:responseJxJybare} \\
    & G^{\rm r}_{J_x, J_x} (\omega) = \left(1 + \left(\frac{\omega_{\rm p}}{\tilde \omega}\right)^2 \frac{\tilde \omega}{2} D_0^{\rm s} (\omega)\right)^2 \frac{G^{\rm r, 0}_{J_x, J_x}  (\omega)}{1 - V_{\rm ind}(\omega) \frac{m}{2 e^2 \tilde \omega N} G^{\rm r, 0}_{J_x, J_x}(\omega)} + \frac{N e^2}{m} \left(\frac{\omega_{\rm p}}{\tilde \omega}\right)^2 \frac{\tilde \omega}{2} D_0^{\rm s} (\omega) \,, \label{eq:responseJxJxbare}\\
    & G^{\rm r}_{J_y, J_y} (\omega) = \frac{G^{\rm r, 0}_{J_y, J_y}(\omega) + V_{\rm ind}(\omega) \frac{m}{2 e^2 \tilde \omega N} \left(G^{\rm r, 0}_{J_x, J_y}(\omega) G^{\rm r, 0}_{J_y, J_x}(\omega) - G^{\rm r, 0}_{J_x, J_x}(\omega) G^{\rm r, 0}_{J_y, J_y}(\omega) \right)}{1 - V_{\rm ind}(\omega) \frac{m}{2 e^2 \tilde \omega N} G^{\rm r, 0}_{J_x, J_x}(\omega)}  \,.
    \label{eq:responseJyJybare}
\end{align}
\end{widetext}
Obtaining explicit expressions for the current response functions is now just a matter of computing the current response functions of the bare electron gas, $G^{\rm r, 0}_{J_r, J_s}$. We refer to the later as bare current response functions and we compute them in the next section.

\subsection{Computing the bare current response functions}

To study the bare 2D electron gas we lean on the reasoning laid out in Ref. \cite{rokaj2023weakened}. There, they study the same 2D electron gas subject to a perpendicular classical magnetic field and coupled to a cavity \eqref{eq:Hqhe}. They show that the cavity only couples to the center-of-mass coordinates. Furthermore, they show that the center-of-mass and relative coordinate sectors of the Hamiltonian commute. This essentially splits the problem into two, the problem of the relative coordinates unmodified by the cavity, and the problem of the center or mass coupled to the cavity. Note from Eqs. \eqref{eq:Jx} and \eqref{eq:Jy} that the currents also only depend on center-of-mass coordinates. Most importantly, the Coulomb interaction between electrons only depends on the relative coordinates. In summary, the full problem of a 2D electron gas coupled to a cavity can be broken into the problem of the relative coordinates, subject to Coulomb interactions but decoupled from the cavity, and the problem of an independent center of mass coordinate, which couples to the cavity and is solely responsible for the conduction properties of the electron gas. Consequently, in order to compute bare current response functions, we only need to study the center-of-mass sector of the matter Hamiltonian
\begin{equation}
    H_{\rm m}^{\rm{cm}} = \frac{\left(\boldsymbol P - e \boldsymbol A_{\rm{ext}} (\boldsymbol R)\right)^2}{2m} \,,
\end{equation}
with $\sqrt{N} \boldsymbol R = \sum_j \boldsymbol r_j$ and $\sqrt{N} \boldsymbol P = \sum_j \boldsymbol p_j$ the center-of-mass position and momentum operators. This is just the Hamiltonian of a single electron under a classical magnetic field, the prototipical example of Landau quantization \cite{tong2016lectures}[Chap. 1]. 
The eigenstates of the system are separable, into plane waves in the $X$ direction and eigenstates of a displaced harmonic oscillator of frequency $\omega_{\rm c}$ along the $Y$ direction
\begin{equation}
\begin{split}
    H_{\rm m}^{\rm{cm}} | K_X, n \rangle & = \omega_{\rm c} \left(c^\dagger c+ \frac{1}{2} \right) | K_X, n \rangle \\
    & = \omega_{\rm c} \left(n + \frac{1}{2}\right) | K_X, n \rangle \,,
\end{split}
\end{equation}
With $P_X | K_X, n \rangle = K_X | K_X, n \rangle$.
Here
\begin{align}
    V = \sqrt{\frac{1}{2 \omega_{\rm c}}} \left(c + c^\dagger\right)  \,, \\
    P_V = i \sqrt{\frac{\omega_{\rm c}}{2}} \left(c^\dagger - c \right) \,,
\end{align}
and $V = \sqrt{m} (Y + K_X l_c^2)$, $P_V = P_Y / \sqrt{m}$, with $l_c = \sqrt{1/(eB)}$.

From the definition in Eq. \eqref{eq:Jwithcavity} (ignoring the contribution from the cavity to get the bare current) we can see that the current only depends on center-of-mass coordinates
\begin{equation}
    \boldsymbol{J} = \sqrt{N} \frac{e}{m} \left(\boldsymbol P - e \boldsymbol A_{\rm ext}(\boldsymbol R) \right) \,.
\end{equation}
After some manipulation, we can express
\begin{align}
    & J_x = \frac{e \omega_{\rm c}}{\sqrt{2m}} \left(\frac{1}{\omega_{\rm c} \sqrt{m}} (P_X - K_X) + \frac{1}{\sqrt{\omega_{\rm c}}} (c + c^\dagger) \right) \,, \\
    & J_y = i \frac{e \sqrt{\omega_{\rm c}}}{\sqrt{2m}}\left(c^\dagger - c\right) \,,
\end{align}
and the matrix elements
\begin{align}
\begin{split}
    \langle K_X', n| J_x | K_X, m \rangle & = \delta_{K_X, K_X'} \frac{e \sqrt{\omega_{\rm c}}}{\sqrt{2m}} (\sqrt{m+1} \delta_{n, m+1}\\
    & \qquad \qquad \qquad \quad \quad + \sqrt{m} \delta_{n, m-1}) \,, \label{eq:matrixelJx}
\end{split}
    \\
\begin{split}
    \langle K_X', n| J_y | K_X, m \rangle & = \delta_{K_X, K_X'} i \frac{e \sqrt{\omega_{\rm c}}}{\sqrt{2m}} (\sqrt{m+1} \delta_{n, m+1} \\
    & \qquad \qquad \qquad \quad \quad - \sqrt{m} \delta_{n, m-1}) \,.
\end{split}
     \label{eq:matrixelJy}
\end{align}
This shows that the current operators are diagonal with respect to the plane waves along the $X$ direction, such that the spectral decomposition of the current response functions can be written simply as a sum over eigenstates of the harmonic oscillator along the $Y$ direction. At zero temperature, this reads
\begin{equation}
    G^{\rm r, 0}_{J_r, J_s}(\omega) = \sum_{n} \left( \frac{\langle 0 | J_r|n\rangle \langle n | J_s | 0 \rangle}{\omega_+ + E_0 - E_n } -  \frac{\langle n | J_r|0\rangle \langle 0 | J_s | n \rangle}{\omega_+ + E_n - E_0} \right) \,. 
     \label{eq:spectraldecompcurrentresponse}
\end{equation}
Note that $\omega_+ = \omega + i \delta$, with $\delta$ a broadening parameter. From Eqs. \eqref{eq:matrixelJx}, \eqref{eq:matrixelJy} and \eqref{eq:spectraldecompcurrentresponse} we obtain the bare current response functions
\begin{align}
\begin{split}
    & G^{\rm r, 0}_{J_x, J_x}(\omega) = G^{\rm r, 0}_{J_y, J_y} (\omega) \\
    & \qquad \qquad \;= - \frac{N e^2 \omega_{\rm c}}{m} \frac{1}{2} \left(\frac{1}{\omega_+ + \omega_{\rm c}} - \frac{1}{\omega_+ - \omega_{\rm c}} \right)
\end{split}
     \,, \label{eq:freeresponseJxJx} \\
    & G^{\rm r, 0}_{J_x, J_y}(\omega) = \frac{N e^2 \omega_{\rm c}}{m} \frac{i}{2} \left(\frac{1}{\omega_+ + \omega_{\rm c}} + \frac{1}{\omega_+ - \omega_{\rm c}} \right) \,,\label{eq:freeresponseJxJy} \\
    & G^{\rm r, 0}_{J_y, J_x}(\omega) = -\frac{N e^2 \omega_{\rm c}}{m} \frac{i}{2} \left(\frac{1}{\omega_+ + \omega_{\rm c}} + \frac{1}{\omega_+ - \omega_{\rm c}} \right) \label{eq:freeresponseJyJx} \,.
\end{align}

\subsection{Closed expressions for the current response functions}

Substituting Eqs. \eqref{eq:freeresponseJxJx}, \eqref{eq:freeresponseJxJy}, \eqref{eq:freeresponseJyJx} and $V_{\rm ind} (\omega) = \omega_{\rm p}^2 D_0^{\rm s}(\omega)$, with $D_0^{\rm s}(\omega)$ the analytic continuation of $D_0^{\rm s}(\omega_m)|_{i\omega_m \to \omega_+}$ \eqref{eq:D0s}, into Eqs. \eqref{eq:responseJxJy}, \eqref{eq:responseJxJx} and \eqref{eq:responseJyJy} yields, after some manipulation,
\begin{align}
    G^{\rm r}_{J_x, J_y} (\omega) = i\frac{N e^2}{m} \frac{\omega_{\rm c} \omega_+\left(\omega_+^2 - \Omega^2\right)}{\left(\omega_+^2 - \tilde \omega^2\right)\left(\omega_+^2 - \omega_{\rm c}^2\right) - \omega_{\rm p}^2 \omega_{\rm c}^2} \,, \label{eq:explicitresponseJxJy}\\
    G^{\rm r}_{J_x, J_x} (\omega) = \frac{N e^2}{m} \frac{(\omega_{\rm c}^2 + \omega_{\rm p}^2)\left(\omega_+^2 - \frac{\Omega^2 \omega_{\rm c}^2}{\omega_{\rm c}^2 + \omega_{\rm p}^2}\right)}{\left(\omega_+^2 - \tilde \omega^2\right)\left(\omega_+^2 - \omega_{\rm c}^2\right) - \omega_{\rm p}^2 \omega_{\rm c}^2} \,, \label{eq:explicitresponseJxJx} \\
    G^{\rm r}_{J_y, J_y} (\omega) = \frac{N e^2}{m} \frac{\omega_{\rm c}^2\left(\omega_+^2 - \Omega^2\right)}{\left(\omega_+^2 - \tilde \omega^2\right)\left(\omega_+^2 - \omega_{\rm c}^2\right) - \omega_{\rm p}^2 \omega_{\rm c}^2} \,. \label{eq:explicitresponseJyJy}
\end{align}
These current response functions have poles at $\omega_+ = \pm \Omega_{\pm}$, with $\Omega_{\pm}$ the frequencies of the Landau polaritons
\begin{equation}
    2 \Omega_\pm = \tilde \omega^2 + \omega_{\rm c}^2 \pm \sqrt{(\tilde \omega^2 - \omega_{\rm c}^2)^2 + 4 \omega_{\rm c}^2 \omega_{\rm p}^2} \,.
\end{equation}
This is all in agreement with the Landau polaritons and current response functions computed in Ref. \cite{rokaj2023weakened}. There, they solve the system exactly, computing the energies and wavefunctions and subsequently the current response functions from the spectral decomposition formula. Therefore, the agreement serves as validation that our effective theory correctly predicts the response functions of a material coupled to a cavity. We also recover the dc conductivities in the long-wavelength limit $\Omega \to 0$, $\delta \to 0$ of Ref. \cite{rokaj2022polaritonic}.

For completeness, let us write down the dc ($\omega \to 0$) conductivities, which are tipically of interest when considering the quantum Hall effect,
\begin{align}
    \sigma_{xy} = \frac{e^2 \nu}{h} \frac{\omega_{\rm c}^2 (\Omega^2 + \delta^2)}{(\Omega_-^2 + \delta^2)(\Omega_+^2 + \delta^2)} \,,\\
    \sigma_{xx} = \sigma_{\rm D} \left(1 - \frac{(\omega_{\rm c}^2 + \omega_{\rm p}^2) \left(\frac{\Omega^2 \omega_{\rm c}^2}{\omega_{\rm c}^2 + \omega_{\rm p}^2} + \delta^2\right)}{(\Omega_-^2 + \delta^2)(\Omega_+^2 + \delta^2)}\right) \,, \\
    \sigma_{yy} = \sigma_{\rm D} \left(1 -  \frac{\omega_{\rm c}^2 (\Omega^2 + \delta^2)}{(\Omega_-^2 + \delta^2)(\Omega_+^2 + \delta^2)}\right) \,.
\end{align}
Note that $\sigma_{\rm D} = e^2 \rho_{\rm{2D}} / (m \delta)$ is the Drude dc conductivity and that in $\sigma_{xy}$ we have introduced the Landau level filling factor $\nu = \rho_{\rm{2D}} h / (e B)$.

It is also interesting to consider the limit of vanishing classical magnetic field $\omega_{\rm c} \to 0$, where the system is just the bare electron gas coupled to a cavity. In this case, we can see from Eqs. \eqref{eq:explicitresponseJxJy}, \eqref{eq:explicitresponseJxJx} and \eqref{eq:explicitresponseJyJy} that only $G^{\rm r}_{J_x, J_x}$ is non-zero. This stems from the fact that all the bare current response functions vanish in this limit, Cf. Eqs \eqref{eq:freeresponseJxJx}, \eqref{eq:freeresponseJxJy} and \eqref{eq:freeresponseJyJx}. Despite this, $G^{\rm r}_{J_x, J_x}$ does not vanish because it contains a term that depends solely on the cavity, i.e. the cavity acts as a current channel in the direction of the light-matter coupling independent of whether the material is a conductor or not. This is evidenced by Eq. \eqref{eq:primitiveresponseJxJx} where the last term solely depends on the free photonic propagator. With this, the optical conductivity reads
\begin{equation}
    \sigma_{xx}(\omega) = \sigma_{\rm D} \frac{i \delta}{\omega_+} \left(1 + \frac{\omega_{\rm p}^2}{\omega_+^2 - \tilde \omega^2} \right) \,.
\end{equation}
This is in agreement with the optical conductivity of the 2D free electron gas computed in Ref. \cite{rokaj2022free}. Once again, there, they solve the system exactly, computing the energies and wavefunctions and subsequenctly the current response functions.

\section{Spin models in a cavity}
\label{spinmodels}

In this section we apply our LRT to a number of spin models. 
We consider these models in the context of magnetic systems, which couple to the magnetic field of the cavity via the Zeeman coupling and as such do not present a $P^2$ term \cite{romanroche2021photon,romanroche2022effective}.
In the particular case of the Dicke model, for the sake of completeness, we will compare the cases with and without the $P^2$ term, $\zeta = 0,1$, since it was historically proposed to describe electric dipoles \cite{dicke1954coherence}, in which case the $P^2$ term is necessary to ensure gauge-invariance \cite{rzazewski1975phase, rzazewski1976remark}.
Thus, with the $P^2$ term the Dicke model describes electric dipoles and without it it describes magnetic dipoles.
Then, we will proceed to consider the rest of the models without the $P^2$ term, assuming that they describe magnetic dipoles.

\subsection{Dicke model}
\label{dicke}

\begin{figure}
    \centering
    \includegraphics[width=\columnwidth]{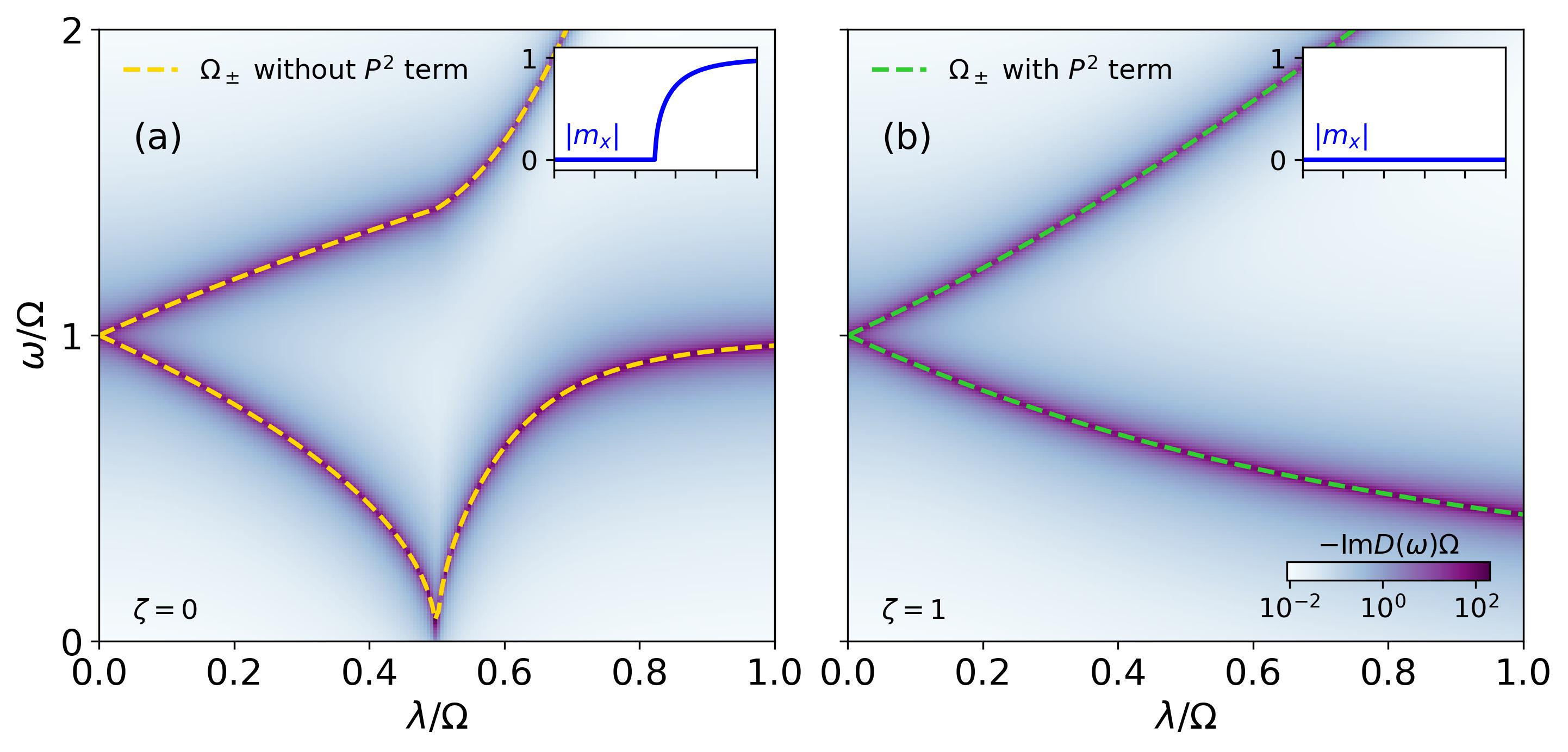}
    \caption{Cavity response, $D$, of the Dicke model as a function of the collective coupling, $\lambda$, without the $P^2$ term, $\zeta=0$, (a) and with the $P^2$ term, $\zeta=1$, (b). 
    The dashed lines correspond to the exact polaritons computed via bosonization. 
    The spins and cavity are resonant, $\omega_z = \Omega$.}
    \label{fig:dicke}
\end{figure}

We apply now our LRT to the paradigmatic Dicke model \cite{dicke1954coherence}, for which $H_{\rm m} = \frac{\omega_z}{2} \sum_j \sigma_j^z$ and $C_x = \sum_j \sigma_j^x$. Here the matter subsystem is just a collection of independent emitters and the only interactions are those mediated by the cavity.
Accordingly
\begin{equation}
    H_{\rm{eff}}^{\rm{MF}} = \frac{\omega_z}{2} \sum_j \sigma_j^z + \frac{2\lambda^2 (\zeta - 1)}{\Omega} m_x \sum_j \sigma_j^x -  \frac{\lambda^2 (\zeta - 1)}{\Omega} m_x^2 \,
    \label{eq:dickeHeff}
\end{equation}
with $m_x = N^{-1} \langle \sum_j \sigma_j^x \rangle$. Note that for $\zeta = 1$, i.e. in the presence of a $P^2$ term, the effective term vanishes and $H_{\rm{eff}}^{\rm{MF}} = H_{\rm m}$. In this case there is no phase transition, as the effective Hamiltonian describes free spins, and $m_x = 0$. For $\zeta=0$, Hamiltonian \eqref{eq:dickeHeff} is non-trivial and can be solved variationally with respect to $m_x$ (See App. \ref{app:meanfieldsinglespin} for details) obtaining, at zero temperature,
\begin{equation}
    m_x = \begin{cases}
        0  & {\rm if} \quad \lambda^2 \leq \frac{1}{4} \omega_z \Omega \,, \\
        \sqrt{1 - \left(\frac{\Omega \omega_z}{4 \lambda^2}\right)^2}  & {\rm if} \quad \lambda^2 > \frac{1}{4} \omega_z \Omega \,.
    \end{cases}
\end{equation}
We see that for $\zeta = 0$ the model exhibits a phase transition between a paramagnetic phase with $m_x = 0$ and a ferromagnetic phase (also termed superradiant when the emphasis is put on the cavity) with $m_x \neq 0$.
In any case, the response function reads 
\begin{equation}
    \tilde \chi_{xx, 0} (\omega) = -\frac{\omega_z^2}{\varepsilon^2} \frac{2 \varepsilon}{\omega_+^2 - \varepsilon^2} \,,
\end{equation}
with $\varepsilon^2 = \omega_z^2 + \left(\frac{4 \lambda^2 (1 - \zeta)}{\Omega} m_x\right)^2$. It depends on the value of $m_x$ and thus on the phase. 

The photonic propagator, $D(\omega)$, which exhibits poles at the resonant frequencies of the hybrid system, is shown in Fig. \ref{fig:dicke}. 
In Fig. \ref{fig:dicke}(a) we see that for $\zeta=0$ there is a complete softening of the lower polariton at the critical coupling $\lambda^2 = \omega_z \Omega / 4$, signaling the phase transition.
In contrast, in Fig. \ref{fig:dicke}(b) we see that for $\zeta=1$ the softening is partial and no phase transition occurs.

The exact polaritons can be obtained with a bosonization of the Dicke model in the thermodynamic limit [See App. \ref{app:twooscillator} for details]. 
In the case of $\zeta = 0$, which corresponds to the standard Dicke model, these read \cite{emary2003chaos}
\begin{widetext}
    \begin{equation}
    2 \Omega_\pm^2 = \begin{cases}
        \omega_z^2 + \Omega^2 \pm \sqrt{\left(\omega_z^2 - \Omega^2\right)^2 + 16 \lambda^2 \omega_z \Omega} & {\rm if} \quad \lambda^2 \leq \frac{1}{4} \omega_z \Omega \,, \\
        \omega_z^2/\mu^2 + \Omega^2 \pm \sqrt{\left(\omega_z^2/\mu^2 - \Omega^2\right)^2 + 4 \omega_z^2 \Omega^2} & {\rm if} \quad \lambda^2 > \frac{1}{4} \omega_z \Omega \,,
    \end{cases}
    \label{eq:dickepolaritons}
\end{equation}
\end{widetext}
with $\mu = \omega_z \Omega / (4 \lambda^2)$. In the case of $\zeta = 1$, the same analysis is possible, except in this case there is only a paramagnetic phase due to the presence of the $P^2$ term. In the bosonization, the $P^2$ term affecting the spin becomes an $A^2$ term of the corresponding oscillator. It can be eliminated with a Bogoliubov transform, yielding a renormalized frequency $\tilde \omega_z^2 = \omega_z(\omega_z + 4 \lambda^2/\Omega)$ and coupling $\tilde \lambda = \lambda \left(1 + 4 \lambda^2/(\omega_z \Omega) \right)^{-1/4}$. Then, the polaritons are just those of the standard Dicke model in the normal phase (top case of Eq. \eqref{eq:dickepolaritons}), with the substitution $\omega_z \to \tilde \omega_z$ and $\lambda \to \tilde \lambda$. For both $\zeta = 0$ and $\zeta = 1$ we find perfect agreement between the exact Dicke polaritons and the poles of the propagator computed with our linear response theory.

\subsection{Longitudinal Dicke-LMG model}
\label{dickeLMG}

We consider now the Lipskin-Meshkov-Glick (LMG) model with longitudinal coupling to the cavity as a marginal generalization of the Dicke model. In contrast with the Dicke model, this one incorporates intrinsic all-to-all interactions in the matter subsystem \cite{herreraromero2022critical, herreraromero2024phase}.

The full Hamiltonian reads
\begin{equation}
\begin{split}
    H =& \frac{\omega_z}{2} \sum_j \sigma_j^z - \frac{J}{N} \sum_{ij} \sigma_i^x \sigma_j^x \\
    & + \Omega a^\dagger a + g (a + a^\dagger) \sum_j \sigma_j^x\,.
\end{split}
    \label{eq:HLMG}
\end{equation}
Interestingly, the collective nature of the intrinsic interaction allows two alternative treatments within the linear response theory. The first option is to treat it like we would any other intrinsic interaction: bundle it together with the field term to form the matter Hamiltonian, $H_{\rm m} = \frac{\omega_z}{2} \sum_j \sigma_j^z - \frac{J}{N} \sum_{ij} \sigma_i^x \sigma_j^x$, and then follow the linear response theory as described in Sec. \ref{LRT}. The second option exploits the collective nature of the intrinsic interaction and treats this term not within $H_{\rm m}$ but explicitly within the linear response theory, analogously to the $P^2$ term, as it is, in fact, just a negative $P^2$ term [Cf. Eqs. \eqref{eq:Hstart} and \eqref{eq:HLMG}]. We will follow this second option for its simplicity, since, as we show below, the resulting $H_{\rm{eff}}^{\rm{MF}}$ corresponds to free spins and is identical in structure to the one from the Dicke model \eqref{eq:dickeHeff}. Accordingly, we have
\begin{equation}
    V_{\rm ind}(\omega) = \frac{2 \lambda^2}{\Omega} \frac{\Omega^2}{\omega^2 - \Omega^2} - 2J \,,  
\end{equation}
and
\begin{equation}
     H_{\rm{eff}}^{\rm{MF}} = \frac{\omega_z}{2} \sum_j \sigma_j^z - 2 J_{\rm eff} m_x \sum_j \sigma_j^x + J_{\rm eff} m_x^2 \,,
\end{equation}
with $m_x = N^{-1} \langle \sum_j \sigma_j^x \rangle$ and $J_{\rm eff} = \frac{\lambda^2}{\Omega} + J$. Solving the Hamiltonian variationally with respect to $m_x$ (See App. \ref{app:meanfieldsinglespin} for details) yields, at zero temperature,
\begin{equation}
    m_x = \begin{cases}
        0  & {\rm if} \quad \omega_z \geq 4 J_{\rm eff} \,, \\
        \sqrt{1 - \left(\frac{\omega_z}{4 J_{\rm eff}}\right)^2}  & {\rm if} \quad \omega_z < 4 J_{\rm eff} \,.
    \end{cases}
\end{equation}
Like the Dicke model (for $\zeta = 0$), the LMG model exhibits a phase transition between a paramagnetic phase with $m_x = 0$ and a ferromagnetic phase with $m_x \neq 0$. The response function reads (See App. \ref{app:meanfieldsinglespin} for details)
\begin{equation}
    \tilde \chi_{xx, 0} (\omega) = -\frac{\omega_z^2}{\varepsilon^2} \frac{2 \varepsilon}{\omega_+^2 - \varepsilon^2} \,,
\end{equation}
with $\varepsilon^2 = \omega_z^2 + \left(4 J_{\rm eff} m_x\right)^2$. It depends on the value of $m_x$ and thus on the phase.

\begin{figure}
    \centering
    \includegraphics[width=\columnwidth]{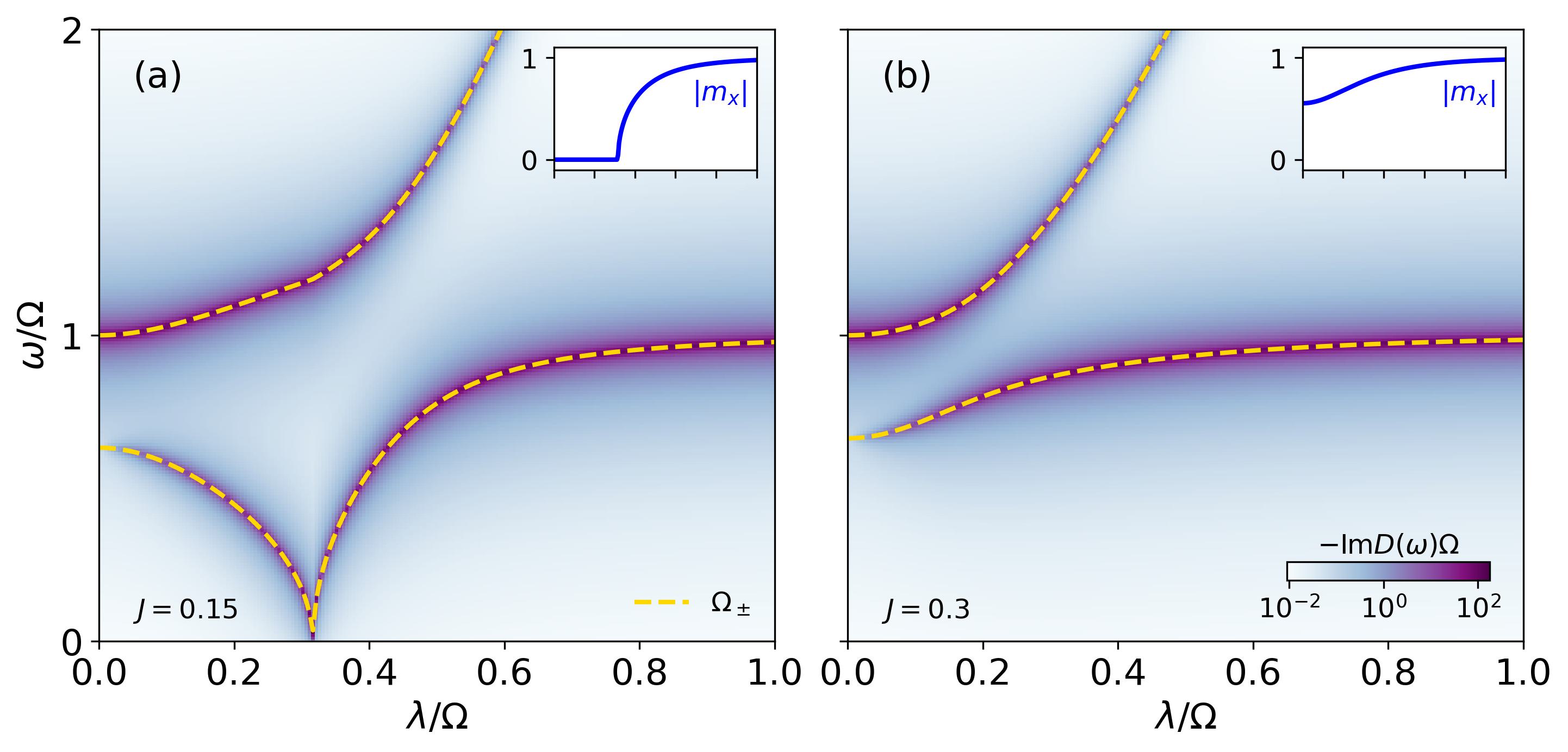}
    \caption{Cavity response, $D$, of the LMG model with longitudinal coupling to the cavity as a function of the collective coupling, $\lambda$, for different values of the intrinsic interaction, $J$. The dashed lines correspond to the exact polaritons computed via bosonization. The top right insets show the magnetization. The transverse field is set to $\omega_z = \Omega$.}
    \label{fig:LMG}
\end{figure}
In Fig. \ref{fig:LMG} we show the photonic propagator, $D(\omega)$, which exhibits poles at the resonant frequencies of the hybrid system, as a function of the collective coupling. In Fig. \ref{fig:LMG}(a), for $J < \omega_z /4$, the lower mode goes to zero at finite coupling, signaling the phase transition. The critical point is displaced to a lower coupling, $\lambda$, with respect to the Dicke model due to the synergy between the intrinsic and effective interactions. In Fig. \ref{fig:LMG}(b), for $J > \omega_z / 4$ the system is always on the ordered phase independently of the value of $\lambda$. In both cases there is a splitting at zero coupling, $\lambda = 0$, due to the intrinsic interactions. In Fig. \ref{fig:LMG} we also overlay the exact polaritons obtained with a bosonization of the longitudinal Dicke-LMG model with longitudinal coupling to the cavity in the thermodynamic limit. It is is a generalization of the bosonization of the Dicke model, developed in Ref. \cite{emary2003chaos}, to account for the interaction term [See App. \ref{app:twooscillator} for details]. It is also a particular case of the polaritons obtained in Ref. \cite{herreraromero2024phase} for the anisotropic Dicke model with collective matter interactions. The resulting LMG polaritons are
\begin{widetext}
    \begin{equation}
    2 \Omega_\pm^2 = \begin{cases}
        \tilde \omega_z^2 + \Omega^2 \pm \sqrt{\left(\tilde \omega_z^2 - \Omega^2\right)^2 + 16 \tilde \lambda^2 \omega_z \Omega} & {\rm if} \quad \omega_z \geq 4 J_{\rm eff} \,, \\
        \omega_z^2/\tilde \mu^2 - 4 \tilde \mu J \omega_z + \Omega^2 \pm \sqrt{\left(\omega_z^2/\tilde \mu^2 - 4 \tilde \mu J \omega_z + \Omega^2\right)^2 + 16 \lambda^2 \tilde \mu \omega_z \Omega} & {\rm if} \quad \omega_z < 4 J_{\rm eff} \,,
    \end{cases}
    \label{eq:LMGpolaritons}
\end{equation}
\end{widetext}
with $\tilde \omega_z^2 = \omega_z(\omega_z - 4J)$, $\tilde \lambda = \lambda (1 - 4J/\omega_z)^{-1/4}$ and $\tilde \mu = \omega_z / (4 J_{\rm eff})$. We find perfect agreement between the exact LMG polaritons and the poles of the propagator computed with our linear response theory.

This model describes recent experiments in rare earth quantum Ising systems, such as LiHoF$_4$ \cite{libersky2021direct, mckenzie2022theory}. These experiments aim to observe how the lowest excitation mode softens as it approaches the ferromagnetic transition. We can understand that Eq. \eqref{eq:HLMG} is equivalent to a Dicke model where $\lambda$ becomes $\sqrt{\lambda^2 + J\Omega}$. Therefore, the softening of the mode can be viewed as the behavior of the lower polariton in the Dicke model.

\subsection{Transverse Dicke-LMG model}
\label{sec:transverseLMG}

Thus far we have only considered models where the intrinsic and cavity-mediated interactions are synergistic, here we consider the transverse Dicke-LMG model: an LMG model with transverse coupling to the cavity \cite{herreraromero2022critical, herreraromero2024phase}, named in analogy to the (transverse) Dicke-Ising model \cite{lee2004firstorder, gammelmark2011phase, zhang2014quantum, cortese2017polariton, rohn2020ising, schellengerger2024almost, langheld2024quantum}. The full Hamiltonian reads
\begin{equation}
\begin{split}
    H =& \frac{\omega_x}{2} \sum_j \sigma_j^x + \frac{\omega_z}{2} \sum_j \sigma_j^z - \frac{J}{N} \sum_{ij} \sigma_i^z \sigma_j^z \\
    & + \Omega a^\dagger a + g\left(a + a^\dagger \right) \sum_j \sigma_j^x \,.
\end{split}
\end{equation}
For vanishing longitudinal and transverse fields, $\omega_z$ and $\omega_x$ respectively, the transverse Dicke-LMG model has a $\mathbb Z_2 \times \mathbb Z_2$ symmetry. The first symmetry corresponds to a spin flip, $\sigma_j^z \to -\sigma_j^z$, and in the bare LMG model it is spontaneously broken in a second order phase transition from a paramagnetic to a ferromagnetic phase. The second symmetry corresponds to a simultaneous cavity-field and spin flip, $a \to -a$ and $\sigma_j^x \to -\sigma_j^x$, and in the bare Dicke model it is spontaneously broken in a second order phase transition from a normal to a superradiant phase. As we show in the following, we find that the combination of the two symmetries gives rise to a first-order phase transition in the transverse Dicke-LMG model between two symmetry-broken phases: a ferromagnetic normal phase for large $J$ and a paramagnetic superradiant phase for large $\lambda^2/\Omega$. Switching on the external field has the effect of explicitly breaking the corresponding symmetry, preventing the spontaneous symmetry breaking that characterizes a phase transition. We find that if only one external field is switched on, the phase transition is demoted from first to second order. If the two external fields are switched on, the phase transition is eliminated completely.

In order to solve the model, we will consider the intrinsic collective interactions explicitly within the linear response theory.  However, unlike in the case of the longitudinal Dicke-LMG model, in this case the cavity-mediated and intrinsic interactions act on different axes, i.e. they couple different collective operators: $C_x = \sum_j \sigma_j^x$ and $C_z = \sum_j \sigma_j^z$ respectively. Accordingly, a slight generalization of the linear response theory presented in Sec. \ref{LRT} is required as there are now two induced interaction terms within the effective action
\begin{align}
    & V_{{\rm ind}, x}(\omega) = \frac{2 \lambda^2}{\Omega} \frac{\Omega^2}{\omega_+^2 - \Omega^2} \,, \\
    & V_{{\rm ind}, z}(\omega) = -2J \label{eq:Vindz} \,.
\end{align}
The system can be solved with the multimode generalization of the linear response theory developed in App. \ref{app:multimode}, since from Eq. \eqref{eq:Sindmultimode} onward it is irrelevant whether the different induced interaction terms all stem from actual cavity modes or if some stem from cavity modes and others from intrinsic collective interactions. The expression for the photonic propagator remains unchanged, see Eq. \ref{eq:relpropphtonmatter}, as it solely determined by the operator that couples to the cavity. In contrast, matter correlators have new contributions afforded by the new interaction channel, $V_{\rm{ind}, z}$. We find
\begin{widetext}
    \begin{equation}
    \chi_{xx} = \frac{\tilde \chi_{xx, 0} + V_{{\rm ind}, z} \det(\tilde \chi_0) }{1 + V_{{\rm ind}, x} \tilde \chi_{xx, 0} + V_{{\rm ind}, z} \tilde \chi_{zz, 0} + V_{{\rm ind}, x}  V_{{\rm ind}, z} \det(\tilde \chi_0)} \,,
\end{equation}
\end{widetext}
with $(\tilde \chi_0)_{rs} = \tilde \chi_{rs, 0}$. The mean-field effective Hamiltonian reads
\begin{equation}
    H_{\rm{eff}}^{\rm{MF}} = \frac{\tilde \omega_x}{2} \sum_j \sigma_j^x +  \frac{\tilde \omega_z}{2} \sum_j \sigma_j^z + \frac{N \lambda^2}{\Omega} m_x^2 + N J m_z^2 \,,
    \label{eq:HeffMFLMGtransverse}
\end{equation}
with $\tilde \omega_x = \omega_x - 4 \frac{\lambda^2}{\Omega}m_x$ and $\tilde \omega_z = \omega_z - 4Jm_z$. Solving variationaly with respect to $m_x$ and $m_z$ (See App. \ref{app:meanfieldsinglespin} for details) allows us to compute the equilibrium values of $m_x$ and $m_z$ numerically and subsequently the response functions $\tilde \chi_{xx, 0}$, $\tilde \chi_{xz, 0}$, $\tilde \chi_{zx, 0}$ and $\tilde \chi_{zz, 0}$, which depend on the values of $m_x$ and $m_z$. 

\begin{figure}
    \centering
    \includegraphics[width=\columnwidth]{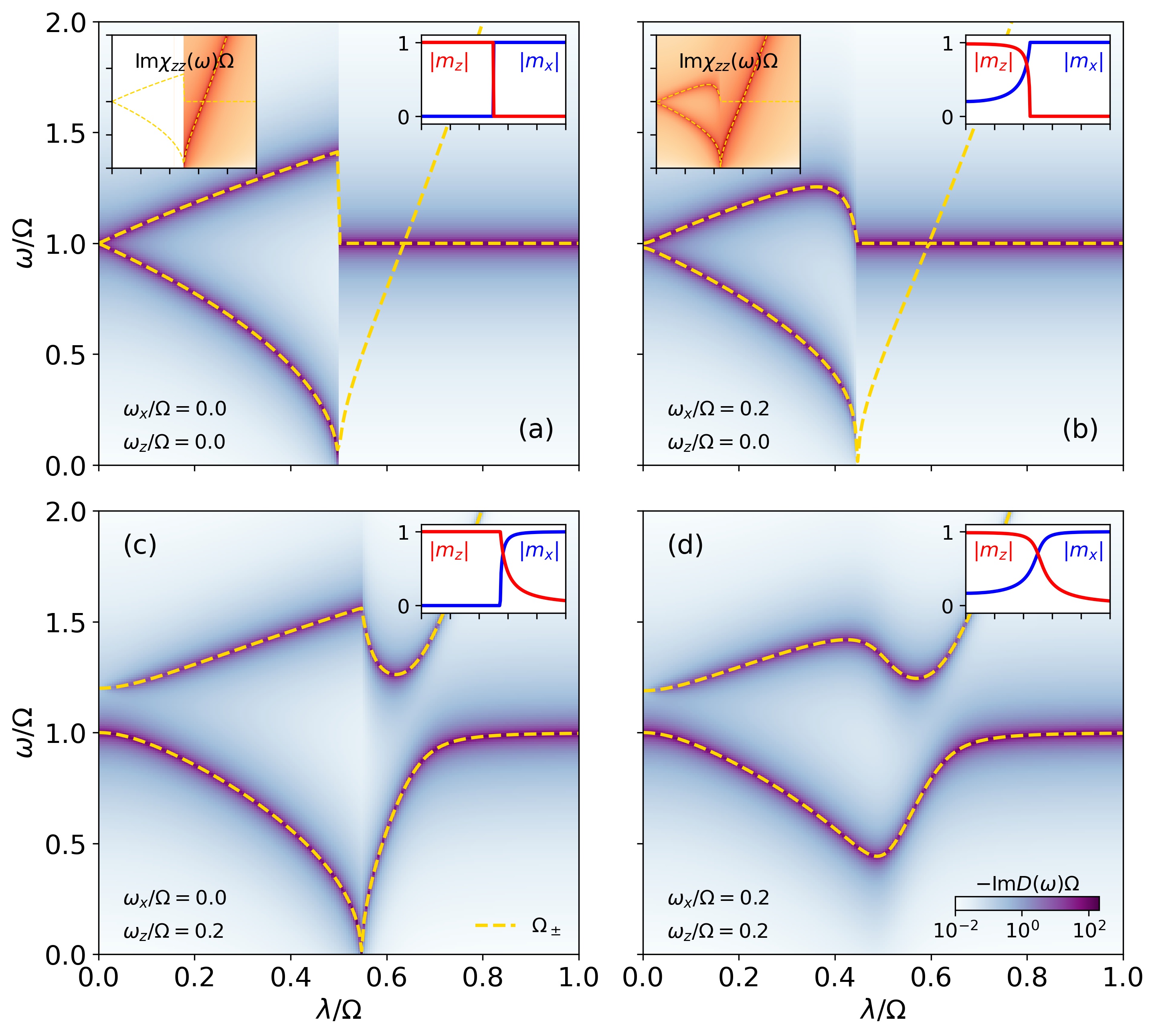}
    \caption{Cavity response, $D$, of the Dicke-LMG model (LMG model with transverse coupling to the cavity) as a function of the collective coupling, $\lambda$, for different values of the longitudinal and transverse fields, $\omega_z$ and $\omega_x$ respectively. The top left insets show the matter response $\chi_{zz}$. The dashed lines correspond to the exact polaritons computed via bosonization. The top right insets show the magnetizations. The intrinsic interaction is set to $4J = \Omega$.}
    \label{fig:transverse-LMG}
\end{figure}

In Fig. \ref{fig:transverse-LMG} we show the photonic propagator, $D(\omega)$, the matter response function, $\chi_{zz}(\omega)$, (left inset) and the magnetizations, $m_x$ and $m_z$, (right inset) as functions of the collective coupling, $\lambda$, in four different scenarios of external fields values. The intrinsic interaction is set so that at zero coupling the cavity and the spins are resonant: $4J = \Omega$. In the case of vanishing $\omega_x$ and $\omega_z$, Fig. \ref{fig:transverse-LMG}(a), we observe a first order phase transition, evidenced by the discontinuous behaviour in the order parameters of the two ordered phases. The photonic propagator has a pole that goes to zero at the critical point, signaling the gap closing. We include $\chi_{zz}$ in the left inset for completeness because the photonic propagator depends on $\chi_{xx}$ [Cf. Eq. \eqref{eq:relpropphtonmatter}] and the system becomes unresponsive to probing along the $x$ direction in the superradiant phase. This is caused by the fact that the intrinsic interaction is the only term in the Hamiltonian containing $\sigma^z$ operators. Due to its collective nature, its effect is switched off (mean-field behaviour) in the superradiant phase where $m_z = 0$ [Cf. Eq. \eqref{eq:HeffMFLMGtransverse}]. The system is then fully ordered along $x$ and with no off-diagonal terms it becomes unresponsive to probing with $\sigma^x$. Accordingly, in the superradiant phase the photonic propagator only shows a pole at the cavity frequency. To witness the other excitation of the system we look at $\chi_{zz}$ which shows a pole that emerges from zero at the critical point, signaling the gap reopening. The combination of $D$ and $\chi_{zz}$ provides a complete picture of the excitations of the system.  In Fig. \ref{fig:transverse-LMG}(b) we see that switching on the transverse field, $\omega_x$, eliminates the superradiant symmetry. The phase transition becomes of second order, with $m_z$ the sole order parameter. Like in the previous case, the absence of non mean-field terms containing $\sigma^z$ in the Hamiltonian makes the system unresponsive to $\sigma_x$ in the superradiant phase. Thus, we also include $\chi_{zz}$ to witness the gap reopening. Alternatively, in Fig. \ref{fig:transverse-LMG}(c) we see that switching on the longitudinal field, $\omega_z$, eliminates the ferromagnetic symmetry. The phase transition becomes of second order, with $m_x$ as the sole order parameter. In this case, the presence of the longitudinal field endows the photonic propagator with visibility of all the excitations, showcasing the full gap closing and reopening at the critical point. Finally in Fig. \ref{fig:transverse-LMG}(d) we switch on both the longitudinal and transverse fields. All symmetries are now explicitely broken and the phase transition gives way to a smooth crossover of $m_z$ and $m_x$ dominant regimes for small and large $\lambda$ respectively. The gap remains finite at all times.

In Fig. \ref{fig:transverse-LMG} we also overlay the exact polaritons obtained with a bosonization of the transverse Dicke-LMG model in the thermodynamic limit. It is a generalization of the bosonization in Refs. \cite{emary2003chaos, herreraromero2024phase}[See App. \ref{app:twooscillator} for details]. In contrast with previous examples, the resulting polaritons do not have a closed expression for $\omega_x \neq 0$ because the superradiant symmetry is explicitely broken in this case. Nevertheless, we find perfect agreement between the exact polaritons and the poles of the photonic and matter response functions computed with our linear response theory.

\subsection{Heisenberg ferromagnet coupled to a cavity}

We finish the section by commenting on the paradigmatic Heisenberg model. In particular we consider a ferromagnetic Heisenberg model on a $3D$ lattice, such that the model is ordered below a critical temperature,
\begin{equation}
    H_{\rm m} = -J \sum_{\langle ij \rangle} \boldsymbol \sigma_i \boldsymbol \sigma_{j} + \frac{\omega_z}{2} \sum_j \sigma_j^z \,.
\end{equation}
Here $\sum_{\langle ij \rangle}$ is the sum over nearest-neighbor pairs.
The ground state of the model corresponds to a symmetry broken ferromagnetic state. For vanishing field, $\omega_z = 0$, the spins will be magnetized in a random direction of space, spontaneously breaking the $SO(3)$ symmetry of the model. The addition of a field demotes the symmetry to $SO(2)$ in the plane perpendicular to the field, setting the magnetization direction along the field. It also opens a gap between the spin-wave band and the ground state, such that the zero momentum spin-wave has an energy gap with respect to the ground state of $\omega_z$. The coupling to a cavity in a direction perpendicular to the classical field will demote the remaining $SO(2)$ symmetry to $\mathbb Z_2$ in the cavity field direction. The cavity coupling, $C_x = \sum_j \sigma_j^x$, will compete with the classical field in setting the magnetization direction. The corresponding effective Hamiltonian reads
\begin{equation}
\begin{split}
    H_{\rm{eff}}^{\rm{MF}} =& \frac{\omega_z}{2} \sum_j \sigma_j^z -J \sum_{\langle ij \rangle} \boldsymbol \sigma_i \boldsymbol \sigma_{j} \\
    &- \frac{2\lambda^2 }{\Omega} m_x \sum_j \sigma_j^x +  \frac{\lambda^2}{\Omega} m_x^2 \,
\end{split}
    \label{eq:HeffMFheisenberg}
\end{equation}
If we assume that the model is ordered, the interaction term contributes an energy $-J z$ with $z$ the coordination number, independently of the value of $m_x$. Thus, the variational Hamiltonian is equivalent to that of a Dicke model [Cf. Sec. \ref{dicke}.]. The behaviour of the magnetization, $m_x$, is also inherited from the Dicke model. For a subcritical coupling, $\lambda < \sqrt{\omega_z \Omega}/2$, the model is ordered along $z$. When the coupling reaches the critical point, the magnetization starts to turn towards the $x$ axis in a second order phase transition. 

The response function $\tilde \chi_{xx, 0}$ corresponds to the excitation of the zero-momentum magnon, with a pole at $\omega^2 = \omega_z^2 + (4 \lambda^2 m_x / \Omega)^2$. Therefore, our LRT predicts the formation of polaritons of this zero-momentum magnon with frequencies given by Eq. \ref{eq:dickepolaritons}, following the physics of the Dicke model.

\section{Conclusions}

In this paper, we have presented a linear response theory for materials collectively coupled to a cavity. We have employed two different approaches. The first relies on a path integral formulation of the partition function and a saddle point expansion of the partition function that is truncated exactly in the thermodynamic limit. The second is more direct and relies on formulating the equations of motion for the response functions. These are truncated to second order in light-matter fluctuations and solved, yielding equivalent expressions for the response functions to the first approach. This provides a posteriori validation for the truncation of the equations of motion, which confirms the validity of a mean-field decoupling of the light-matter interaction in cavity QED materials. We obtain exact expressions for the response functions of the hybrid system in terms of the bare response functions of the cavity and the material.

The theory has been demonstrated in several systems, showcasing its applicability. In the Quantum Hall effect we have recovered the optical response and Landau polaritons computed previously in Refs. \cite{rokaj2022polaritonic, rokaj2023weakened}. In magnetic systems, we have started from the non-interacting Dicke model and progressively complicated the model by adding collective intrinsic interactions to yield the longitudinal Dicke-LMG model and then changing the cavity coupling direction to induce a competition between intrinsic and light-matter interactions in the transverse Dicke-LMG model. Finally, we have explored the effect of the cavity in an ordered Heisenberg ferromagnet. Our predictions have been validated against other analytical results.

Even if the focus of the paper has been the linear response theory, the study of the different spin models has provided some interesting independent physical insights. In particular that first order phase transitions arise in hybrid systems with competing orders, such as the transverse Dicke-LMG model. In this sense, our results also point to the fact that the robustness of this first order phase transition depends on the nature of the intrinsic material interactions. A transverse field demotes the phase transition to second order for the transverse Dicke-LMG model (collective interactions) but not for for the (transverse) Dicke-Ising model (nearest-neighbor interactions) \cite{romanroche2024linear,lee2004firstorder,gammelmark2011phase,langheld2024quantum}. We underline that this has been witnessed through the lens of the photonic response, which is relevant for its experimental implications. Cavity transmission is a common probing mechanism for magnetic materials inductively coupled to a cavity \cite{rollano2022high, macneill2019gigahertz, zollitsch2023probing, mergenthaler2017strong, martinezlosa2023measuring}. 

Although not explored here, the computation of finite-size corrections is possible. In the path-integral approach this would be achieved by considering higher order terms in the saddle-point expansion. Alternatively, one could truncate the equations of motion at higher order, to include three- or higher-order correlators~\cite{queisser2023hierarchy, gomezleon2016hierarchy, gomezleon2017hierarchy}. This would be relevant in the study of topological systems, where boundary effects are crucial \cite{perezgonzalez2022topology, dmytruk2022controlling,perezgonzalez2023lightmatter}. Also, real time dynamics could be easily studied with the framework of the equations of motion~\cite{gomezleon2019spin,rincon2024mumax3cqed}.

\section*{Acknowledgements}
We acknowledge discussions with Katharina Lenk, Martin Eckstein, Andreas Wipf,  Yuto Ashida, Luis Martín-Moreno, Fernando Falceto and José G. Esteve. Despite the enormous impediments in spending the budget, the authors must acknowledge funding from the grant
TED2021-131447B-C21 funded by MCIN/AEI/10.13039/501100011033 and the EU
‘NextGenerationEU’/PRTR. We also acknoledge grant CEX2023-001286-S financed by MICIU/AEI /10.13039/501100011033, the Gobierno de Aragón (Grant E09-17R Q-MAD), Quantum Spain and the CSIC Quantum
Technologies Platform PTI-001. J. R-R acknowledges support from the Ministry of Universities of the Spanish Government through the grant FPU2020-07231.
\appendix

\section{Linear response theory for a multimode cavity}
\label{app:multimode}
Consider the Hamiltonian for a material coupled collectively to a multimode cavity
\begin{equation}
\begin{split}
    H =& H_{\rm m} + \sum_k \Omega_k a_k^\dagger a_k \\
    &+ \sum_k g_k \left(a_k + a_k^\dagger\right) C_{k} + \zeta \sum_k \frac{g_k^2}{\Omega_k} C_k^2\,. 
\end{split}
    \label{eq:Hstartmultimode}
\end{equation}
The corresponding action reads
\begin{equation}
\begin{split}
    S =&  S_{\rm m} + \sum_k \int_\tau \bar a_k(\tau) (\partial_\tau - \Omega_k) a_k(\tau) \\
    & + \sum_k g_k \int_\tau (a_k(\tau) + \bar a_k(\tau)) C_{k}(\tau) \\
    & + \zeta \sum_k \frac{g_k^2}{\Omega_k} \int_\tau C_{k}^2(\tau) \,.
\end{split}
\label{eq:ogactionmultimode}
\end{equation}
After partial integration over the cavity field, the induced action reads
\begin{equation}
    S_{\rm ind} = \frac{1}{2} \sum_k \int_{\tau, \tau'} C_{k}(\tau) \frac{1}{N} V_{\rm ind, k}(\tau - \tau') C_{k}(\tau') \,,
    \label{eq:Sindmultimode}
\end{equation}
with 
%
\begin{equation}
    V_{\rm ind, k} (\omega_m) = 2 \lambda_k^2 \frac{\Omega_k^2 (\zeta - 1) + \zeta \omega_m^2}{\Omega_k (\omega_m^2 + \Omega_k^2)} \,.
\end{equation}
The induced interaction can be decoupled with a Hubbard-Stratonovich transformation that introduces an auxiliary scalar field for each mode, $\varphi_k$:
\begin{widetext}
\begin{equation}
    e^{-S_{\mathrm{ind}}}=\frac{1}{Z_\varphi} \oint_{\varphi} e^{-\frac{1}{2} N \sum_k \int_{\tau, \tau'}  \varphi_k(\tau) V_{\mathrm{ind}, k}^{-1}(\tau-\tau^{\prime}) \varphi_k(\tau^{\prime})-i \sum_k \int_\tau  \varphi_k(\tau) C_{k}(\tau)}\,.
    \label{newHSmultimode}
\end{equation}
\end{widetext}
We define the propagator of the auxiliary fields as
\begin{equation}
    W_{kp}(\tau) = \langle \varphi_k(\tau) \varphi_p(0) \rangle^{\rm c} \,.
\end{equation}
The generating functional for bare connected correlation functions of the matter coupling operators, $C_{k}$, is $\mathcal G_{\rm m}^0[\xi_k] = - N^{-1} \log Z_{\rm m}[\xi_k]$ with
\begin{equation}
    Z_{\rm m}[\xi_k] = \oint_{c} e^{-(S_{\rm m} + i \sum_k \int_\tau \xi_k(\tau) C_{k}(\tau))} \,,    
    \label{eq:generatingfuncfreemattermultimode}
\end{equation}
such that 
\begin{equation}
\begin{split}
    \frac{\delta}{\delta \xi_{k_1}(\tau_1)} \cdots & \frac{\delta}{\delta \xi_{k_n}(\tau_n)} \mathcal G_{\rm m}^0[\xi_k] \Bigr|_{\xi_k = 0} = \\
    & = - \frac{(-i)^n}{N}  \langle C_{k_1}(\tau_1) \cdots C_{k_n}(\tau_n)\rangle^{\rm c}_{\rm m} \\
    & \equiv \chi_{k_1 \ldots k_n, 0}^{(n)}(\tau_1, \ldots, \tau_n) \,. 
\end{split}
\end{equation}
In the following we will denote $\chi_{kp, 0}^{(2)} \equiv \chi_{kp, 0}$. With this, after partial integration over the matter degrees of freedom, we can write
\begin{equation}
    Z = \oint_{\varphi_k} e^{-N f[\varphi_k]}  \,,
\end{equation}
with
\begin{equation}
    f[\varphi_k] =  \frac{1}{2} \sum_k \int_{\tau, \tau'} \varphi_k(\tau) V_{\mathrm{ind}, k}^{-1}(\tau-\tau^{\prime}) \varphi_k(\tau^{\prime} ) + \mathcal G_{\rm m}^0[\varphi_k] \,.
    \label{eq:fmultimode}
\end{equation}
Like in the single-mode case, we apply a saddle-point approximation for large $N$. This requires that the number of modes, $M$, be finite, i.e. that $\lim_{N \to \infty} M/N \to 0$. Otherwise higher order terms in the saddle-point expansion might not actually be negligible. The condition that the functional derivatives of $f$ with respect to $\varphi_k$ vanish
%
\begin{equation}
    \varphi_{k, \rm sp}(\omega_m) =  -\frac{i}{N} V_{\rm ind, k}(\omega_m) \langle C_{k} (\omega_m) \rangle_{\varphi_{k, \rm sp}}
    \label{eq:saddlepointconditionmultimode}
\end{equation}
formally defines $\varphi_{k, \rm sp}$. Assuming a constant solution for $\varphi_{k, \rm sp}$, we find
\begin{equation}
    \varphi_{k, \rm sp} = -\frac{i}{N} V_{\rm{ ind}, k}(\omega_m = 0) \langle C_k\rangle_{\varphi_{k, \rm sp}} \,.
    \label{eq:constphi0conditionmultimode}
\end{equation}
This tells us that $\varphi_{k, \rm sp}$ is self-consistently proportional to $\langle C_x \rangle_{\varphi_{k, \rm sp}}$, i.e. to the expectation value of $C_k$ for the bare matter subject to fields $i \varphi_{k, \rm{sp}} =\frac{1}{N} V_{\rm ind, k}(\omega_m = 0) \langle C_k \rangle$. This is precisely the self-consistent condition that arises from computing $\langle C_k\rangle$ from the mean-field Hamiltonian 
\begin{equation}
\begin{split}
    H_{\rm eff}^{\rm MF} =&  H_{\rm m} + \sum_k \frac{2\lambda_k^2 (\zeta - 1)}{N \Omega_k} \langle C_k \rangle C_k \\
    & - \sum_k \frac{\lambda_k^2 (\zeta - 1)}{N \Omega_k} \langle C_k \rangle^2 \,.
    \label{eq:Heffmultimode}
\end{split}
\end{equation}
This is the mean-field theory of the effective Hamiltonian that arises from taking the static limit in $V_{\rm ind, k}$ in the effective action \cite{romanroche2022effective}.

Expanding $f$ around the saddle-point yields
\begin{equation}
     (N W_{0})_{kp}^{-1} = \delta_{kp} V_{\rm{ind}, k}^{-1} + \tilde \chi_{kp,0}\,,
     \label{eq:W0identitymultimode}
\end{equation}
with 
\begin{equation}
\begin{split}
    \tilde \chi_{kp, 0} (\tau_1, \tau_2) = & \frac{\delta}{\delta \varphi_{k, \rm sp}(\tau_1)}\frac{\delta}{\delta \varphi_{p, \rm sp}(\tau_2)} \mathcal G_{\rm m}^0 \\
     = & \frac{1}{N} \langle C_k(\tau) C_p(\tau')\rangle_{\varphi_{k, \rm{sp}}}^{\rm c} \,.
\end{split}
    \label{eq:chi0tildemultimode}
\end{equation}
For $N \to \infty$ one can safely truncate the saddle-point expansion to second order, which implies that $W = W_0$. 

We can now obtain relations between the auxiliary field propagator, $W_{kp}$, the photonic propagator, $D_k$, and matter response functions. Let us define the generating functional for photonic connected correlators by introducing a complex fields in Eq. \eqref{eq:ogactionmultimode}: $\mathcal G_{\rm ph} [\eta_k, \bar \eta_k] = - \log Z[\eta_k, \bar \eta_k]$, with
\begin{equation}
    Z[\eta_k, \bar \eta_k] = \oint_{a_k, \bar a_k, c} e^{-(S + \sum_k \int_\tau (a_k(\tau) \bar \eta_k(\tau) + \bar a_k(\tau) \eta_k(\tau))} \,.
\end{equation}
After partial integration of the cavity fields, this yields
\begin{equation}
\begin{split}
    Z [\eta_k, \bar \eta_k] = \oint_{c} \exp &- \Bigl( S_{\rm m} + \zeta \sum_k\frac{g_k^2}{\Omega_k} \int_\tau C_{k}^2(\tau) \\
    & + \sum_k \int_{\tau, \tau'} \bar m_k(\tau) D_k(\tau - \tau') m_k(\tau) \Bigr) \,,
\end{split}
\end{equation}
with $m_k(\tau) = \eta_k(\tau) + g_k C_{k}(\tau)$. With this
%
\begin{equation}
    D_k(\omega_m) = D_{k, 0}(\omega_m) - \lambda_k^2 D_{k, 0}(\omega_m) \chi_{kk}(\omega_m) D_{k, 0}(\omega_m) \,.
    \label{eq:relpropphtonmattermultimode}
\end{equation}
Likewise, we can define the generating functional for matter connected correlators: $\mathcal G_{\rm m}[\xi_k] = - N^{-1} \log Z[\xi_k]$ with
\begin{equation}
    Z[\xi_k] = \oint_{c, \varphi_k} e^{-(S_{\rm eff} + i \sum_k \int_\tau \xi_k(\tau) C_{k}(\tau))} \,.
    \label{eq:generatingfuncmatter}
\end{equation}
After partial integration over the cavity fields, this yields
\begin{equation}
    Z[\xi_k] = \oint_{\varphi_k} e^{-N f[\varphi_k, \xi_k]} \,,
\end{equation}
with
\begin{equation}
\begin{multlined}
    f[\varphi_k, \xi_k] = \\
    \frac{1}{2} \sum_k \int_{\tau, \tau'} \varphi_k(\tau) V_{\mathrm{ind}, k}^{-1}(\tau-\tau^{\prime}) \varphi_k(\tau^{\prime} ) + \mathcal G_{\rm m}^0[\varphi_k + \xi_k] \,.
\end{multlined}
\end{equation}
Then, a second order expansion of $f[\varphi_k, \xi_k]$ around the saddle point yields
\begin{widetext}
\begin{equation}
\begin{split}
    f[\varphi_k, \xi_k] - f[\varphi_{k, \rm{sp}}, 0] = \frac{1}{2} \sum_{k, p} \int_{\tau, \tau'}  \Bigl( &\delta \varphi_k(\tau) (NW)_{kp}^{-1}(\tau - \tau')\delta \varphi_p(\tau') + \xi_{k}(\tau) \tilde \chi_{k p, 0}(\tau - \tau') \xi_{p}(\tau') \\
    & + \, \delta\varphi_k(\tau) \tilde \chi_{k p, 0}(\tau - \tau') \xi_{p} (\tau') + \xi_{k} (\tau) \tilde \chi_{k p, 0}(\tau - \tau') \delta \varphi_p(\tau) \Bigr) \,.
\end{split}
\end{equation}
The functional integral over the auxiliary-field displacements $\delta \varphi_k$ is just an $M$-dimensional Gaussian integral that we can perform, yielding
\begin{equation}
    G_{\rm m}[\xi_k] = {\rm cst.} + \frac{1}{2} \sum_{k, p} \int_{\tau, \tau'} \xi_k(\tau) \left( \tilde \chi_{kp, 0}(\tau - \tau') - \sum_{k', p'} \int_{u, v} \tilde \chi_{k k', 0}(\tau - u) N W_{k' p'} (u - v) \tilde \chi_{p' p, 0}(v - \tau') \right) \xi_p(\tau') \,.
\end{equation}
With this
\begin{equation}
    \chi_{kp}(\omega_m) = \tilde \chi_{kp, 0}(\omega_m) - \sum_{k', p'} \tilde \chi_{k k', 0}(\omega_m) N W_{k' p'}(\omega_m) \tilde \chi_{p' p, 0}(\omega_m) \,.
\end{equation}
Note that obtaining final expressions for $\chi_{kp}(\omega_m)$ requires inverting the matrix $(N W)^{-1}$ that is defined in Eq. \eqref{eq:W0identitymultimode}.
\end{widetext}

\section{Equivalence between the photonic propagator in the path-integral and the equations-of-motion formulations}
\label{app:equivalence}
In the equations-of-motion formulation, the photonic propagator is given by Eq. \eqref{eq:Dfromeom}.
Here, we show that after some manipulations, this expression is equivalent to Eqs. \eqref{eq:relpropphtonmatter} and \eqref{eq:chixaidentity} obtained from the path-integral formulation.

First we rewrite Eq. \eqref{eq:Dfromeom} as
\begin{equation}
    D(\omega) = \frac{D_0(\omega) + \lambda^2 \frac{2 \zeta( \omega_+ + \Omega) - \Omega}{\Omega (\omega_+^2 - \Omega^2)} \tilde \chi_{xx, 0}(\omega)}{1 + 2\lambda^2 \frac{\zeta(\omega_+^2 - \Omega^2) + \Omega^2}{\Omega (\omega_+^2 - \Omega^2)} \tilde \chi_{xx, 0}(\omega)}
\end{equation}
Then, we note that
\begin{equation}
    2\lambda^2 \frac{\zeta(\omega_+^2 - \Omega^2) + \Omega^2}{\Omega} = V_{\rm ind}(\omega) \,,
\end{equation}
and similarly
\begin{equation}
    \lambda^2 \frac{2 \zeta( \omega_+ + \Omega) - \Omega}{\Omega} = D_0(\omega) \left(V_{\rm ind}(\omega) - \lambda^2 D_0(\omega)\right) \,.
\end{equation}
With this
\begin{align}
    D(\omega) &= \frac{D_0(\omega) + D_0(\omega)\left(V_{\rm ind}(\omega) - \lambda^2 D_0(\omega) \right) \tilde \chi_{xx, 0}(\omega)}{1 + V_{\rm ind}(\omega) \tilde \chi_{xx, 0}} \notag \\
    & =\frac{D_0(\omega) \left(1 + V_{\rm ind}(\omega) \tilde \chi_{xx, 0}(\omega) \right) - \lambda^2 D_0^2(\omega)}{1 + V_{\rm ind}(\omega) \tilde \chi_{xx, 0}} \notag \\
    & =D_0(\omega) - \lambda^2 D_0(\omega) \frac{\tilde \chi_{xx, 0}(\omega)}{1 + V_{\rm ind}(\omega) \tilde \chi_{xx, 0}(\omega)} D_0(\omega) \,,
\end{align}
which is precisely the analytic continuation $i \omega_m \to \omega_+$ of the combination of Eqs. \eqref{eq:relpropphtonmatter} and \eqref{eq:chixaidentity}.

\section{Implicit expressions for the current response functions in the quantum Hall effect}
\label{app:currentcorrelatorsqhe}
\subsubsection{Computing mixed light-matter correlations}
To compute the mixed Green functions of Eqs. \eqref{eq:mixed1}, \eqref{eq:mixed2} and \eqref{eq:mixed3} following our section \ref{LRT}, we formulate the following generating functional for mixed connected correlators: $\mathcal G_{\rm mix} [\eta, \xi_x, \xi_y] = - \log Z[\eta, \xi_x, \xi_y]$, with
\begin{equation}
\begin{split}
    Z[\eta, \xi_x, \xi_y] = \oint_{a, \bar a, c} \exp - \Bigl(&  S + \int_\tau \eta(\tau) (a(\tau) + \bar a(\tau)) \\
    & + \sum_{r} \int_\tau \xi_r (\tau) C_r(\tau) \Bigr) \,.
\end{split}
\end{equation}
After partial integration of the cavity fields, this yields
\begin{equation}
\begin{split}
     Z[\eta, \xi_x, \xi_y] = \oint_{c}\exp - \Bigl( & S_{\rm m} + \int_{\tau, \tau'} m(\tau) D_0(\tau - \tau') m(\tau) \\
    & + \sum_{r} \int_\tau \xi_r (\tau) C_r(\tau) \Bigr) \,,
\end{split}
\end{equation}
with $m(\tau) = \eta(\tau) + g C_x(\tau)$. With this
\begin{equation}
\begin{split}
    G^{\rm t}_{\gamma(a + a^\dagger), C_r}(\tau - \tau') & =  -\gamma \frac{\delta}{\delta \eta(\tau)} \frac{\delta}{\delta \xi_r (\tau')} \mathcal G_{\rm mix}  [\eta, \xi_x, \xi_y] \\
    & = \gamma \int_u D_0^{\rm s} (\tau - u) N g \chi_{x r}(u - \tau') \,,
\end{split}
\end{equation}
or in Matsubara frequency space
\begin{equation}
    G^{\rm t}_{\gamma(a + a^\dagger), C_r}(\omega_m) = -\gamma D_0^{\rm s}(\omega_m) N g \chi_{xr}(\omega_m) \,.
\end{equation}
Here $D_0^s(\tau) = D_0(\tau) + D_0(-\tau)$ is the symmetrized free photon propagator
\begin{equation}
    D_0^{\rm s}(\omega_m) = \frac{1}{i \omega_m - \tilde \omega} - \frac{1}{i \omega_m + \tilde \omega} \,.
    \label{eq:D0s}
\end{equation}
Likewise
\begin{equation}
    G^{\rm t}_{C_x, \gamma(a + a^\dagger)}(\omega_m) = -\gamma D_0^{\rm s}(\omega_m) N g \chi_{x x}(\omega_m)
\end{equation}
and
\begin{equation}
\begin{split}
    G^{\rm t}_{\gamma(a + a^\dagger), \gamma(a + a^\dagger)}(\omega_m) = & \gamma^2 D_0^{\rm s}(\omega_m) \\
    & - \gamma^2 D_0^{\rm s}(\omega_m) \lambda^2 \chi_{x x}(\omega_m) D_0^{\rm s}(\omega_m) \,.
\end{split}
\end{equation}
With these we obtain, after analytic continuation $i \omega_m \to \omega_+ = \omega + i \delta$,
\begin{equation}
\begin{split}
    G^{\rm r}_{J_x, J_y} (\omega) &= -N \frac{2 e^2 \tilde \omega}{m} \left(1 + \gamma g D_0^{\rm s}(\omega) \right) \chi_{x y}(\omega) \,, \\
    &= -N \frac{2 e^2 \tilde \omega}{m} \left(1 + \left(\frac{\omega_{\rm p}}{\tilde \omega}\right)^2 \frac{\tilde \omega}{2} D_0^{\rm s} (\omega) \right) \chi_{x y}(\omega) \,,
\end{split}
\end{equation}
and similarly
\begin{widetext}
\begin{align}
    & G^{\rm r}_{J_x, J_x} (\omega) = - N \frac{2 e^2 \tilde \omega}{m} \left(\left(1 + \left(\frac{\omega_{\rm p}}{\tilde \omega}\right)^2 \frac{\tilde \omega}{2} D_0^{\rm s}  (\omega)\right)^2 \chi_{x x}(\omega) - \frac{1}{2 \tilde \omega} \left(\frac{\omega_{\rm p}}{\tilde \omega}\right)^2 \frac{\tilde \omega}{2} D_0^{\rm s} (\omega) \right) \,, \label{eq:primitiveresponseJxJx}\\
    & G^{\rm r}_{J_y, J_y} (\omega) = - N \frac{2 e^2 \tilde \omega}{m} \chi_{y y}(\omega) \,.
\end{align}
We can relate $\chi_{xx}$, $\chi_{xy}$ and $\chi_{yy}$ to $\tilde \chi_{xx, 0}$, $\tilde \chi_{xy, 0}$ and $\tilde \chi_{yy, 0}$ through Eqs. \eqref{eq:chixaidentity} and \eqref{eq:chiyyidentity}, obtaining
\begin{equation}
    G^{\rm r}_{J_x, J_y} (\omega) = \left(1 + \left(\frac{\omega_{\rm p}}{\tilde \omega}\right)^2 \frac{\tilde \omega}{2} D_0^{\rm s} (\omega)\right) \frac{\tilde G^{\rm r, 0}_{J_x, J_y}  (\omega)}{1 - V_{\rm ind}(\omega) \frac{m}{2 e^2 \tilde \omega N} \tilde G^{\rm r, 0}_{J_x, J_x}(\omega)} \,,
    \label{eq:responseJxJy}
\end{equation}
with $\tilde G^{\rm r, 0}_{J_r, J_s} = -N \tilde \chi_{rs, 0}$. Likewise
\begin{align}
    & G^{\rm r}_{J_x, J_x} (\omega) = \left(1 + \left(\frac{\omega_{\rm p}}{\tilde \omega}\right)^2 \frac{\tilde \omega}{2} D_0^{\rm s} (\omega)\right)^2 \frac{\tilde G^{\rm r, 0}_{J_x, J_x}  (\omega)}{1 - V_{\rm ind}(\omega) \frac{m}{2 e^2 \tilde \omega N} \tilde G^{\rm r, 0}_{J_x, J_x}(\omega)} + \frac{N e^2}{m} \left(\frac{\omega_{\rm p}}{\tilde \omega}\right)^2 \frac{\tilde \omega}{2} D_0^{\rm s} (\omega) \,, \label{eq:responseJxJx}\\
    & G^{\rm r}_{J_y, J_y} (\omega) = \frac{\tilde G^{\rm r, 0}_{J_y, J_y}(\omega) + V_{\rm ind}(\omega) \frac{m}{2 e^2 \tilde \omega N} \left(\tilde G^{\rm r, 0}_{J_x, J_y}(\omega) \tilde G^{\rm r, 0}_{J_y, J_x}(\omega) - \tilde G^{\rm r, 0}_{J_x, J_x}(\omega) \tilde G^{\rm r, 0}_{J_y, J_y}(\omega) \right)}{1 - V_{\rm ind}(\omega) \frac{m}{2 e^2 \tilde \omega N} \tilde G^{\rm r, 0}_{J_x, J_x}(\omega)}  \,.
    \label{eq:responseJyJy}
\end{align}
\end{widetext}
We will now show that $\tilde \chi_{r s, 0} = \chi_{r s, 0}$ and thus $\tilde G^{\rm r, 0}_{J_r, J_s} = G^{\rm r, 0}_{J_r, J_s}$.

\subsubsection{Proof that $\tilde \chi_{r s, 0} = \chi_{r s, 0}$}
From Eqs. \eqref{eq:Heff} and \eqref{eq:compactHqhe} we see that the mean-field effective Hamiltonian of the electron gas is given by
\begin{equation}
\begin{split}
    H_{\rm eff}^{\rm MF} =&  \sum_j \frac{\left(\boldsymbol p_j + m \omega_{\rm c} y_j \boldsymbol e_x\right)^2}{2m} + \sum_{i > j} V(\boldsymbol r_i - \boldsymbol r_j) \\
    & - 2 \frac{\omega_{\rm p}^2}{\tilde \omega^2} \langle p_{x} + m \omega_{\rm c} y \rangle \sum_j \frac{p_{j,x} + m \omega_{\rm c} y_j}{2m} \\
    & + \rm{cst.} \,,
\end{split}
\end{equation}
with $\langle p_{x} + m \omega_{\rm c} y \rangle = \langle p_{j, x} + m \omega_{\rm c} y_j \rangle \, \forall j$.
Completing squares yields
\begin{equation}
\begin{split}
     H_{\rm eff}^{\rm MF} & = \sum_j \frac{1}{2m} \left(p_{j,x} + m \omega_{\rm c} y_j - \frac{\omega_{\rm p}^2}{\tilde \omega^2}  \langle p_{x} + m \omega_{\rm c} y \rangle\right)^2 \\ 
     & + \sum_j \frac{p_{j,y}^2}{2m}  + \sum_{i > j} V(\boldsymbol r_i - \boldsymbol r_j) + \rm{cst.} \,.
\end{split}
\end{equation}
We can see now that
\begin{equation}
    p_{j,x} + m \omega_{\rm c} y_j - \frac{\omega_{\rm p}^2}{\tilde \omega^2}  \langle p_{x} + m \omega_{\rm c} y \rangle \propto [x_j, H_{\rm eff}^{\rm MF}]
\end{equation}
and since $\langle [x_j, H_{\rm eff}^{\rm MF}] \rangle = 0$ we have
\begin{equation}
    \langle p_{x} + m \omega_{\rm c} y \rangle \left(1 - \frac{\omega_{\rm p}^2}{\Omega^2 + \omega_{\rm p}^2} \right) = 0 \,,
\end{equation}
which implies that $\langle p_{x} + m \omega_{\rm c} y \rangle = 0$ and thus $H_{\rm eff}^{\rm MF} = H_{\rm m}$ and $\tilde \chi_{r s, 0} = \chi_{r s, 0}$. This is a no go theorem, implying that the equilibrium properties of the electron gas are not modified by the cavity. Despite this, the linear response properties are modified, as we show in the following.

This implies that $\tilde G^{\rm r, 0}_{J_r, J_s} = G^{\rm r, 0}_{J_r, J_s}$. With this, Equations \eqref{eq:responseJxJy}, \eqref{eq:responseJxJx} and \eqref{eq:responseJyJy} turn into Eqs. \eqref{eq:responseJxJybare}, \eqref{eq:responseJxJxbare} and \eqref{eq:responseJyJybare} and provide us with an implicit expression of the current response functions of the electron gas coupled to a cavity in terms of the current response functions of the bare electron gas. This culminates the application of our linear response theory.

\section{A spin subject to variational fields}
\label{app:meanfieldsinglespin}
\subsection{Spectrum and response functions of a free spin}
Let us consider the following free spin model
\begin{equation}
    H = \frac{\omega_x}{2} \sigma_x + \frac{\omega_z}{2} \sigma_z \,.
    \label{eq:Hfreespin}
\end{equation}
It is diagonalized by a rotation that defines spin operators along new directions
\begin{align}
    & \sigma_z' = \frac{\omega_z}{\epsilon} \sigma_z -  \frac{\omega_x}{\epsilon} \sigma_x \,, \\
    & \sigma_x' =  \frac{\omega_x}{\epsilon} \sigma_z +  \frac{\omega_z}{\epsilon} \sigma_x \,, 
\end{align}
where $\epsilon^2 = \omega_x^2 + \omega_z^2$.
With these, the Hamiltonian reads
\begin{equation}
    H = \frac{\epsilon}{2} \sigma_z' \,,
\end{equation}
with ground-state energy $E_0 = -\epsilon /2$. 

We compute the response functions from their spectral decomposition, which, at zero temperature, reads
\begin{equation}
    \chi_{rs}(\omega) = -\sum_n \left(\frac{\langle 0 | \sigma_r | n \rangle \langle n | \sigma_s | 0 \rangle}{\omega^+ + E_0 - E_n} - \frac{\langle n | \sigma_r | 0 \rangle \langle 0 | \sigma_s | n \rangle}{\omega^+ + E_n - E_0} \right) \,,
\end{equation}
where the sum is over the eigenstates of the system. With the matrix elements 
\begin{align}
    & \langle n | \sigma_x | 0 \rangle = - \frac{\omega_x}{\epsilon} \,, \\
    & \langle n | \sigma_z | 0 \rangle =  \frac{\omega_z}{\epsilon} \,,
\end{align}
we obtain
\begin{align}
    & \chi_{xx}(\omega) = - \frac{\omega_z^2}{\epsilon^2} \frac{2 \epsilon}{\omega_+^2 - \epsilon^2} \,, \label{eq:chixxfreespin} \\
    & \chi_{xz}(\omega) = \chi_{zx}(\omega) = - \frac{\omega_x \omega_z}{\epsilon^2} \frac{2 \epsilon}{\omega_+^2 - \epsilon^2} \,, \label{eq:chizzfreespin} \\
    & \chi_{zz}(\omega) = - \frac{\omega_x^2}{\epsilon^2} \frac{2 \epsilon}{\omega_+^2 - \epsilon^2} \,. \label{eq:chixzfreespin}
\end{align}

\subsection{Considering variational fields}

There will be situations in which we arrive to a free spin model like the one defined in Eq. \eqref{eq:Hfreespin} from a mean field approximation. As a result we will have a Hamiltonian of the form
\begin{equation}
     H(m_x, m_z) = \frac{\tilde \omega_x}{2} \sigma_x + \frac{\tilde \omega_z}{2} \sigma_z + \Delta(m_x, m_z) \,,
\end{equation}
where now the fields, $\tilde \omega_x$ and $\tilde \omega_z$ depend on mean-field parameters $m_{r \in \{x, z\}} = \langle \sigma_r \rangle$ and there is a constant energy term that also depends on the variational parameters. The Hamiltonian can be diagonalized as explained in the previous section, and the ground state energy and response functions are now functions of $m_x$ and $m_z$. In particular, $E_0(m_x, m_z) = -\epsilon(m_x, m_z) /2 + \Delta(m_x, m_z)$, with $\epsilon^2 = \tilde \omega_x^2 + \tilde \omega_z^2$. The values of $m_x$ and $m_z$ are determined variationaly from a minimization of $E_0(m_x, m_z)$. Depending on the particular problem, this can be done analytically or numerically. 

\section{Bosonization of Dicke-LMG models}
\label{app:twooscillator}

In Ref. \cite{emary2003chaos} the ground-state sector of the Dicke model is solved exactly in the thermodynamic limit. 
A Holstein-Primakov transformation is used to bosonize the collective spin representing the $N$ two-level atoms of the Dicke model.
The resulting model is that of two coupled quantum harmonic oscillators and can be diagonalized with a Bogoliubov transformation to obtain the normal modes of the system.
These are the polaritons of the Dicke model.
To describe the superradiant (symmetry-broken) phase of the Dicke model, a displacement is introduced in the two oscillators prior to invoking the thermodynamic limit, to account for their finite occupation at equilibrium.
Here we extend this technique to the longitudinal and transverse Dicke-LMG models considered in Sec. \ref{spinmodels}. Similar extensions have been developed in Ref. \cite{herreraromero2024phase}.

\subsection{Longitudinal Dicke-LMG model}

The ground-state sector of the Hamiltonian of the longitudinal Dicke-LMG model [See Eq. \eqref{eq:HLMG}] can be written in terms of maximum-total-spin operators as
\begin{equation}
\begin{multlined}
        H = \omega_z S_z - \frac{J}{N}\left(S_+ + S_- \right)^2 + \Omega a^\dagger a \\+ \frac{\lambda}{\sqrt{N}}\left(S_+ + S_- \right)\left(a + a^\dagger\right) \,,
\end{multlined}
\end{equation}
with $S_z$, $S_+$, $S_-$ spin $S=N/2$ operators.
The Holstein-Primakov transformation provides a mapping between bosonic annihilation and creation operators and spin operators
\begin{align}
    &S_z = b^\dagger b - N \,, \\
    &S_+ = \sqrt{N} b^\dagger \sqrt{1 - \frac{b^\dagger b}{N}} \,, \\
    &S_- = \sqrt{N} \sqrt{1 - \frac{b^\dagger b}{N}} b \,,
\end{align}
with $[b, b^\dagger] = 1$. Ignoring constants, the resulting bosonic Hamiltonian reads
\begin{equation}
    \begin{multlined}
        H = \omega_z b^\dagger b - J\left(b^\dagger \sqrt{1 - \frac{b^\dagger b}{N}} + \sqrt{1 - \frac{b^\dagger b}{N}} b \right)^2 \\
        + \Omega a^\dagger a + \lambda \left(b^\dagger \sqrt{1 - \frac{b^\dagger b}{N}} + \sqrt{1 - \frac{b^\dagger b}{N}} b \right)\left(a + a^\dagger\right) \,.
    \end{multlined}
\end{equation}
To allow for the description of both a normal and a symmetry-broken phase, in which the bosonic degrees of freedom acquire macroscopic occupations, we displace the bosonic operators as
\begin{align}
    a^\dagger \to c^\dagger + \sqrt{\alpha} \,, \label{eq:displacementcavity}\\
    b^\dagger \to d^\dagger - \sqrt{\beta} \,. \label{eq:displacementspins}
\end{align}
with $\alpha$ and $\beta$ real parameters whose value we will determine later but which are assumed to scale as $N$.
This is equivalent to assuming that in the symmetry broken phases the two modes acquire a macroscopic occupation.
With these displacements and ignoring constants, the Hamiltonian becomes
\begin{equation}
    \begin{multlined}
        H = \omega_z \left(d^\dagger d - \sqrt{\beta} (d + d^\dagger)\right) \\
        - J \frac{k}{N} \left(d^\dagger \sqrt{\xi} + \sqrt{\xi} d -2 \sqrt{\beta} \sqrt{\xi} \right)^2  \\
        + \Omega \left(c^\dagger c + \sqrt{\alpha} (c + c^\dagger)\right) \\
        + \lambda \sqrt{\frac{k}{N}} \left(d^\dagger \sqrt{\xi} + \sqrt{\xi} d -2 \sqrt{\beta} \sqrt{\xi} \right) \left(c + d^\dagger + 2 \sqrt{\alpha} \right) \,,
    \end{multlined}
\end{equation}
with
\begin{equation}
    \sqrt{\xi} = \sqrt{1 - \frac{d^\dagger d - \sqrt{\beta} (d + d^\dagger)}{k}} \,,
\end{equation}
and $k = N - \beta$.
Taking the thermodynamic limit by expanding the square root $\sqrt{\xi}$ and setting terms with overall powers of $N$ in the denominator to zero we obtain
\begin{equation}
    \begin{multlined}
        H = \left\{\omega_z + \frac{2 \lambda}{k} \sqrt{\frac{\alpha \beta k}{2j}} + 4J \frac{\beta}{N}\right\} d^\dagger d \\
        + \left\{\frac{\lambda}{2k^2} \sqrt{\frac{\alpha \beta k}{N}} (2k + \beta) + J \frac{k}{N} \left(\frac{4 \beta}{k} - 1 \right) \right\} \left(d + d^\dagger\right)^2 \\
        + \Omega c^\dagger c + \frac{2\lambda}{k} \sqrt{\frac{k}{N}} \left(\frac{N}{2} - \beta\right) \left(d + d^\dagger\right)\left(c + c^\dagger\right) \\
        + \Biggl\{ \frac{4 \lambda}{k} \sqrt{\frac{\alpha k}{N}} \left(\frac{N}{2} - \beta\right) - \omega_z \sqrt{\beta} \\
        + \frac{8J \sqrt{\beta}}{N} \left(\frac{N}{2} - \beta\right)\Biggr\} \left(d + d^\dagger \right) \\
        + \left\{2 \lambda \sqrt{\frac{\beta k}{N}} - \Omega \sqrt{\alpha}\right\} \left(c + c^\dagger \right)
    \end{multlined}
\end{equation}
To eliminate the linear terms we impose 
\begin{align}
    & 2 \lambda \sqrt{\frac{\beta k}{N}} - \Omega \sqrt{\alpha} = 0 \,, \\
    & \frac{4 \lambda}{k} \sqrt{\frac{\alpha k}{N}} \left(\frac{N}{2} - \beta\right) - \omega_z \sqrt{\beta} + \frac{8J}{N} \left(\frac{N}{2} - \beta\right) \sqrt{\beta} \,.
\end{align}
These equations admit a trivial solution $\sqrt{\alpha} = \sqrt{\beta} = 0$, which corresponds to the normal phase, and a non-trivial solution for $\omega_z < J_{\rm eff}$
\begin{align}
    & \sqrt{\alpha} = \frac{\lambda}{\Omega} \sqrt{N (1 - \tilde \mu^2)} \,,  \label{eq:nontrivialsolalpha}\\
    & \sqrt{\beta} =  \sqrt{\frac{N}{2} (1 - \tilde \mu)} \label{eq:nontrivialsolbeta}\,,
\end{align}
with $\tilde \mu = \omega_z / (4 J_{\rm eff})$ and $J_{\rm eff} = J + \lambda^2/\Omega$, which corresponds to the symmetry-broken phase.
Accordingly, the Hamiltonian in the normal phase reads
\begin{equation}
    H = \omega_z d^\dagger d - J \left(d + d^\dagger \right)^2 + \Omega c^\dagger c + \lambda \left(d + d^\dagger\right)\left(c + c^\dagger\right) \,,
\end{equation}
and in the symmetry-broken phase
\begin{equation}
    \begin{multlined}
        H = \frac{\omega_z}{2 \tilde \mu} (1 + \tilde \mu) d^\dagger d + \lambda \tilde \mu \sqrt{\frac{2}{1 + \tilde \mu}} \left(d + d^\dagger \right) \left(c + c^\dagger \right)\\
        + \Omega c^\dagger c + \left\{\omega_z \frac{(1 - \tilde \mu)(3 + \tilde \mu)}{8 \tilde \mu (1 + \tilde \mu)} - 2J \frac{\tilde \mu^2}{1 + \tilde \mu} \right\} \left( d + d^\dagger \right)^2  \,.
    \end{multlined}
\end{equation}
These can be diagonalized with a Bogoliubov transformation to obtain the polaritons of Eq. \eqref{eq:LMGpolaritons}.

\subsection{Transverse Dicke-LMG model}

Here we will apply the same method to obtain the polaritons of the transverse Dicke-LMG model. 
To avoid repetition, we will present the main equations and we will discuss only the differences between the two models.

The ground-state sector of the Hamiltonian of the transverse Dicke-LMG model [See Eq. \eqref{eq:HLMG}] can be written in terms of maximum-total-spin operators as
\begin{equation}
\begin{multlined}
    H = \omega_z S_z + \frac{\omega_x}{2} \left(S_+ + S_- \right) - \frac{4J}{N}S_z^2 + \Omega a^\dagger a \\
    + \frac{\lambda}{\sqrt{N}}\left(S_+ + S_- \right)\left(a + a^\dagger\right) \,,
\end{multlined}
\end{equation}
After the Holstein-Primakov trasnformation, the Hamiltonian reads
\begin{equation}
    \begin{multlined}
        H = \omega_z b^\dagger b  -  \frac{4J}{N} \left( b^\dagger b - \frac{N}{2} \right)^2 + \Omega a^\dagger a \\
        + \frac{\omega_x \sqrt{N}}{2} \left(b^\dagger \sqrt{1 - \frac{b^\dagger b}{N}} + \sqrt{1 - \frac{b^\dagger b}{N}} b \right) \\
         + \lambda \left(b^\dagger \sqrt{1 - \frac{b^\dagger b}{N}} + \sqrt{1 - \frac{b^\dagger b}{N}} b \right)\left(a + a^\dagger\right) \,.
    \end{multlined}
\end{equation}
Once again, we introduce the displacements of Eqs. \eqref{eq:displacementcavity} and \eqref{eq:displacementspins}, which yield
\begin{equation}
    \begin{multlined}
        H = \omega_z \left(d^\dagger d - \sqrt{\beta} (d + d^\dagger)\right)  + \Omega \left(c^\dagger c + \sqrt{\alpha} (c + c^\dagger)\right) \\
        + \frac{\omega_x}{2} \sqrt{k} \left(d^\dagger \sqrt{\xi} + \sqrt{\xi} d -2 \sqrt{\beta} \sqrt{\xi} \right)  \\
        -\frac{4J}{N} \left(d^\dagger d - \sqrt{\beta} (d + d^\dagger) + \beta - \frac{N}{2}\right)^2 \\
        + \lambda \sqrt{\frac{k}{N}} \left(d^\dagger \sqrt{\xi} + \sqrt{\xi} d -2 \sqrt{\beta} \sqrt{\xi} \right) \left(c + d^\dagger + 2 \sqrt{\alpha} \right) \,.
    \end{multlined}
\end{equation}
Here we take the thermodynamic limit by expanding the square root $\sqrt{\xi}$ and the square in the first term of the second line and neglecting terms with overall powers of $N$ in the denominator, obtaining
\begin{equation}
    \begin{multlined}
        H = \left\{\omega_z + \frac{2 \lambda}{k} \sqrt{\frac{\alpha \beta k}{2j}} + \frac{8J}{N} \left(\frac{N}{2} - \beta\right)\right\} d^\dagger d \\
        + \left\{\frac{\lambda}{2k^2} \sqrt{\frac{\alpha \beta k}{N}} (2k + \beta) - 4J\frac{\beta}{N}\right\} \left(d + d^\dagger\right)^2 \\
        + \Omega c^\dagger c + \frac{2\lambda}{k} \sqrt{\frac{k}{N}} \left(\frac{N}{2} - \beta\right) \left(d + d^\dagger\right)\left(c + c^\dagger\right) \\
        + \Biggl\{ \frac{4 \lambda}{k} \sqrt{\frac{\alpha k}{N}} \left(\frac{N}{2} - \beta\right) - \omega_z \sqrt{\beta} \\
        - \frac{8J \sqrt{\beta}}{N} \left(\frac{N}{2} - \beta\right)
        + \frac{\omega_x}{\sqrt{k}} \left(\frac{N}{2} - \beta\right)\Biggr\} \left(d + d^\dagger \right) \\
        + \left\{2 \lambda \sqrt{\frac{\beta k}{N}} - \Omega \sqrt{\alpha}\right\} \left(c + c^\dagger \right)
    \end{multlined}
\end{equation}
To eliminate the linear terms we impose 
\begin{align}
    & 2 \lambda \sqrt{\frac{\beta k}{N}} - \Omega \sqrt{\alpha} = 0 \,, \label{eq:linearzero1}\\
    & \begin{multlined}
        \frac{4 \lambda}{k} \sqrt{\frac{\alpha k}{N}} \left(\frac{N}{2} - \beta\right) - \omega_z \sqrt{\beta} \\
        - \frac{8J \sqrt{\beta}}{N} \left(\frac{N}{2} - \beta\right) + \frac{\omega_x}{\sqrt{k}} \left(\frac{N}{2} - \beta\right) = 0 \,.
    \end{multlined}  \label{eq:linearzero2}
\end{align}
In this case, since we are considering a finite longitudinal field $\omega_x$ the symmetry is always explicitly broken, so there is no spontaneous symmetry breaking. 
Accordingly, there is no trivial solution, and in general Eqs. \eqref{eq:linearzero1} and \eqref{eq:linearzero2} do not have an analytical solution. 
Nevertheless, they can be solved numerically, and the resulting Hamiltonian can be diagonalized with a Bogoliubov transform to obtain the polaritons that we show in Fig. \ref{fig:transverse-LMG}.

For the sake of completeness, let us solve analytically the case of $\omega_x = 0$.
In this case, Eqs. \eqref{eq:linearzero1} and \eqref{eq:linearzero2} admit a trivial solution $\sqrt{\alpha} = \sqrt{\beta} = 0$, which corresponds to the normal phase, and a non-trivial solution for $\omega_z < 4(\lambda^2/\Omega - J)$ with the same expression for $\sqrt{\alpha}$ and $\sqrt{\beta}$ as in Eqs. \eqref{eq:nontrivialsolalpha} and \eqref{eq:nontrivialsolbeta} but now with 
\begin{equation}
    \tilde \mu = \frac{\omega_z}{4 \left(\frac{\lambda^2}{\Omega} - J\right)} \,,
\end{equation}
which corresponds to the symmetry-broken phase. The difference in the value of $\tilde \mu$ between this (transverse) case and the longitudinal case showcases the fact that in the longitudinal case there is a synergy between the intrinsic and cavity-mediated interactions whereas in this (transverse) case there is a competition.
Accordingly, the Hamiltonian in the normal phase reads
\begin{equation}
    H = \tilde \omega_z d^\dagger d + \Omega c^\dagger c + \lambda \left(d + d^\dagger\right)\left(c + c^\dagger\right) \,,
\end{equation}
with $\tilde \omega_z = \omega_z + 4J$, and in the symmetry-broken phase
\begin{equation}
    \begin{multlined}
        H =  (1 + \tilde \mu) \left(\frac{\omega_z}{2 \tilde \mu} + 2J\right) d^\dagger d + \Omega c^\dagger c \\
        + \lambda \tilde \mu \sqrt{\frac{2}{1 + \tilde \mu}} \left(d + d^\dagger \right) \left(c + c^\dagger \right) \\
         + \left\{\omega_z \frac{(1 - \tilde \mu)(3 + \tilde \mu)}{8 \tilde \mu (1 + \tilde \mu)} - J \frac{(1 - \tilde \mu) (1 + 3 \tilde \mu)}{2(1 + \tilde \mu)}\right\} \left( d + d^\dagger \right)^2 \,.
    \end{multlined}
\end{equation}
These can be diagonalized with a Bogoliubov transformation to obtain the polaritons.
In the normal phase, the Hamiltonian is that of a Dicke model with a bare spin energy $\tilde \omega_z$.
Thus, the polaritons are the same as for the normal phase in Eq. \eqref{eq:dickepolaritons} with the substitution $\omega_z \to \tilde \omega_z$.
In the symmetry-broken phase, the polaritons can be written formally as 
\begin{equation}
\begin{multlined}
   2 \Omega_\pm^2 = \omega_{\rm A}^2 + 4  \omega_{\rm A} \omega_{\rm B} + \Omega^2 \\
   \pm \sqrt{\left(\omega_{\rm A}^2 + 4  \omega_{\rm A} \omega_{\rm B} - \Omega^2\right)^2 + 16 \omega_{\rm A} \omega_{\rm C}^2 \Omega} \,,
\end{multlined}
\end{equation}
with
\begin{align}
    & \omega_{\rm A} = (1 + \tilde \mu) \left(\frac{\omega_z}{2 \tilde \mu} + 2J\right) \,, \\
    & \omega_{\rm B} = \omega_z \frac{(1 - \tilde \mu)(3 + \tilde \mu)}{8 \tilde \mu (1 + \tilde \mu)} - J \frac{(1 - \tilde \mu) (1 + 3 \tilde \mu)}{2(1 + \tilde \mu)} \,, \\
    & \omega_{\rm C} = \lambda \tilde \mu \sqrt{\frac{2}{1 + \tilde \mu}} \,.
\end{align}

\bibliography{main.bib}

\end{document}